\documentclass[hidelinks,11pt]{article}

\usepackage{amssymb,amsmath,amsfonts,eurosym,geometry,ulem,graphicx,caption,color,setspace,sectsty,comment,footmisc,caption,natbib,pdflscape,subfigure,array,hyperref, subcaption, algpseudocode, algorithm}

\usepackage[raggedrightboxes]{ragged2e}

\newcolumntype{L}[1]{>{\raggedright\let\newline\\arraybackslash\hspace{0pt}}m{#1}}
\newcolumntype{C}[1]{>{\centering\let\newline\\arraybackslash\hspace{0pt}}m{#1}}
\newcolumntype{R}[1]{>{\raggedleft\let\newline\\arraybackslash\hspace{0pt}}m{#1}}

\usepackage{mathtools}
\usepackage{dsfont}
\usepackage{bbold}
\begin{document}

\begin{titlepage}
\title{Non-linear Phillips Curve for India: Evidence from Explainable Machine Learning\thanks{We are extremely grateful to Hans Amman (Editor-in-Chief) and two anonymous reviewers for their insightful comments and suggestions on the paper. We would also like to thank Dr.Tanujit Chakraborty (Sorbonne University Abu Dhabi, SAFIR, Abu Dhabi, United Arab Emirates.), Abhishek Kumar, Monishankar Bishnu, Satyanshu Kumar, Utkarsh Kumar and seminar participants at Indian Statistical Institute, New Delhi for their helpful feedback and discussion. The views expressed in the paper are those of authors and do not necessarily reflect the views of the institutions to which they belong. All authors contributed equally to this research. }}
\author{Bhanu Pratap\thanks{Reserve Bank of India, Mumbai, India. Email: bhanupratap@rbi.org.in} \and Amit Pawar\thanks{Reserve Bank of India, Mumbai, India. Email: amitp@rbi.org.in. (corresponding author)} \and Shovon Sengupta\thanks{Fidelity Investments, Boston, U.S.A. Email: shovon.sengupta@fmr.com. }}
\date{}

\maketitle
\begin{abstract}
The conventional, linear Phillips curve model -- while useful for policymaking -- falls short in terms of forecasting power amidst structural breaks and inherent non-linearities. This paper addresses these shortcomings by applying machine learning (ML) methods within a New Keynesian Phillips Curve framework to forecast and explain headline inflation in India, a large emerging economy. Our forecasting analysis suggests that ML-based methods provide large gains in forecasting accuracy over standard linear models. Further, using \textit{explainable} ML techniques, we empirically show that the Phillips curve relationship in India is a highly non-linear process which is efficiently captured by ML models. While headline inflation is found to be strongly influenced by inflation expectations followed by past inflation and output gap, the relationship exhibits non-linearities in the form of thresholds and interaction effects between covariates. Supply shocks, except rainfall, seem to have a marginal impact on headline inflation. ML models, therefore, not only enhance forecast accuracy but also help uncover complex, non-linear relationships in the data in a flexible manner. \\
\vspace{0in}
\vspace{0in}

\bigskip
\end{abstract}
\setcounter{page}{0}
\thispagestyle{empty}
\end{titlepage}
\pagebreak \newpage

\doublespacing

\section{Introduction} \label{sec:introduction}

Accurate inflation forecasts are of prime importance to policymakers, especially central banks that are entrusted with the responsibility of managing monetary policy in an economy (\citealp{bernanke_2003}). A foundational framework within the literature on inflation dynamics is the Phillips Curve (PC) model. The Phillips Curve posits a short-term trade-off between inflation and a measure of economic slack, typically proxied by unemployment rate, such that higher inflation is associated with lower slack in the economy and vice-versa. The earliest empirical validation of this relationship, based on wage inflation and unemployment rate was provided by \cite{phillips1958relation} for the United Kingdom. Since then, the Phillips Curve framework has undergone significant theoretical advancements, culminating in the development of the micro-founded New Keynesian Phillips Curve (NKPC) \citep{taylor1980aggregate, CALVO1983383,gali1999inflation} as the workhorse model for inflation analysis. Despite its theoretical appeal, the practical application of the NKPC for inflation modelling and forecasting—particularly within central banks—has been fraught with challenges. 

Such difficulties stem from structural breaks, state dependencies, and intrinsic nonlinearities in the relationship between inflation and its fundamental determinants, complicating its empirical validity and predictive performance (see \citealp{cristini2021nonlinear}). These issues are particularly pronounced in case of emerging market economies (EMEs), like India, characterized by rapid structural transformations, evolving labor market dynamics, and inflationary pressures influenced by both domestic and global shocks. Unlike advanced economies with relatively stable inflation dynamics, inflation trajectory in EMEs tend to be shaped by factors such as supply-side constraints, weather shocks leading to food price fluctuations, global oil prices and evolving monetary and fiscal policy frameworks. The recent Covid-19 pandemic further added large-scale volatility in the data adding to the existing challenges for inflation forecasting \citep{lenza2020estimate, bobeica2023covid, carriero2024addressing}. 

Traditionally, such issues have been resolved through non-linear econometric models such as threshold and regime-switching models. The recent integration of machine learning (ML) and econometrics, however, has revolutionized the manner in which these challenges can be addressed (see \citealp{varian2014big, mullainathan2017machine}). Machine learning, a subset of artificial intelligence (AI), leverages advanced statistical techniques to discern patterns and relationships in large and complex datasets. Within economics, ML techniques are now increasingly being employed to forecast macroeconomic and financial indicators. These models have delivered newer insights to stakeholders due to their greater predictive power and the ability to capture nonlinear relationships in the data (\citealp{athey2019machine, coulombe2022machine}). Yet ML methods, despite their proven track record of providing superior forecasts, have most often been critiqued on grounds of being a \emph{black-box} approach. This does not bode well from a public policy or business perspective wherein identification and quantification of a ‘cause-and-effect’ attains utmost importance (see \citealp{doshi2017interpretable, miller2017explainable}). 

To address this critique, emerging research has opened the doors for \emph{explainable} ML techniques -- novel methods designed to open the black box -- to act as a pivotal bridge between the forecasting power of ML models and the interpretability demanded by practitioners.\footnote{While there is no strict definition of interpretation, it can be defined as the ability to present or explain a given model in terms understandable by humans.} To implement such an explanation for an ML model, one must rely on an algorithm that relates the feature (input) values of the given model with its prediction to generate \emph{local} interpretation i.e., explain an instance-level prediction. Similarly, one can also generate \emph{global} interpretation of model output at the level of the sample dataset. By allowing a transparent framework to analyze the impact of each variable on model predictions and understand nonlinear relationship between various variables, such techniques equip both model developers and end users to derive better insights from the ML model.\footnote{In simple terms, \emph{interpretability} is about developing an understanding of the \emph{cause-and-effect} within an artificial intelligence (AI) system. In the literature, it refers to the degree in which one can measure or estimate the outcome of a model given an input, understand how the predictions may change with changes in input data or algorithmic parameters and finally understand when the model can go wrong. \emph{Explainability}, on the other hand, refers primarily to the process that helps us understand in a human-readable form how and why the model came up with the predictive outcome(s).} This paper, therefore, seeks to bring together macroeconomics and machine learning by applying a ML-based forecasting and explanation framework to estimate a nonlinear Phillips curve for India. 

Like other EMEs, inflation dynamics in India have been influenced by a host of domestic and global factors. Given its high dependence on oil imports and a large share of agriculture sector, crude oil prices and climate-related factors, such as rainfall, affected inflation in India. Thus, for much of its history, inflation in India has remained high, often times in double digits, lowering the macroeconomic stability of the country. Thereafter, in October 2016, India formally adopted a flexible inflation targeting (FIT) framework to guide its monetary policy in targeting a publicly announced inflation target of 4 per cent (+/- 2 per cent). This marked a significant policy shift in one of the most populous and fastest growing economies in the world.\footnote{The RBI Act was amended in March 2016 to institutionalize the Flexible Inflation Targeting (FIT) regime and the first meeting of the six-member Monetary Policy Committee (MPC) was held in October 2016. See \citet{chakravarty_2020}, \citet{dua_2023} and \citet{ghate_ahmed_2023} for more information on the monetary policy landscape in India and the recent transition to FIT regime.} However, after the adoption of the FIT framework, inflation in India returned to lower levels and has remained low since then \citep{eichengreen2024inflation}. Emerging evidence has highlighted the role of credible monetary policy and effective supply management in contributing to keeping inflation low and stable \citep{kishorandpratap2023}. 

\begin{figure}[ht]
 \caption{How does Economic Slack affect Inflation?}
    \centering
    \includegraphics[height = 6cm]{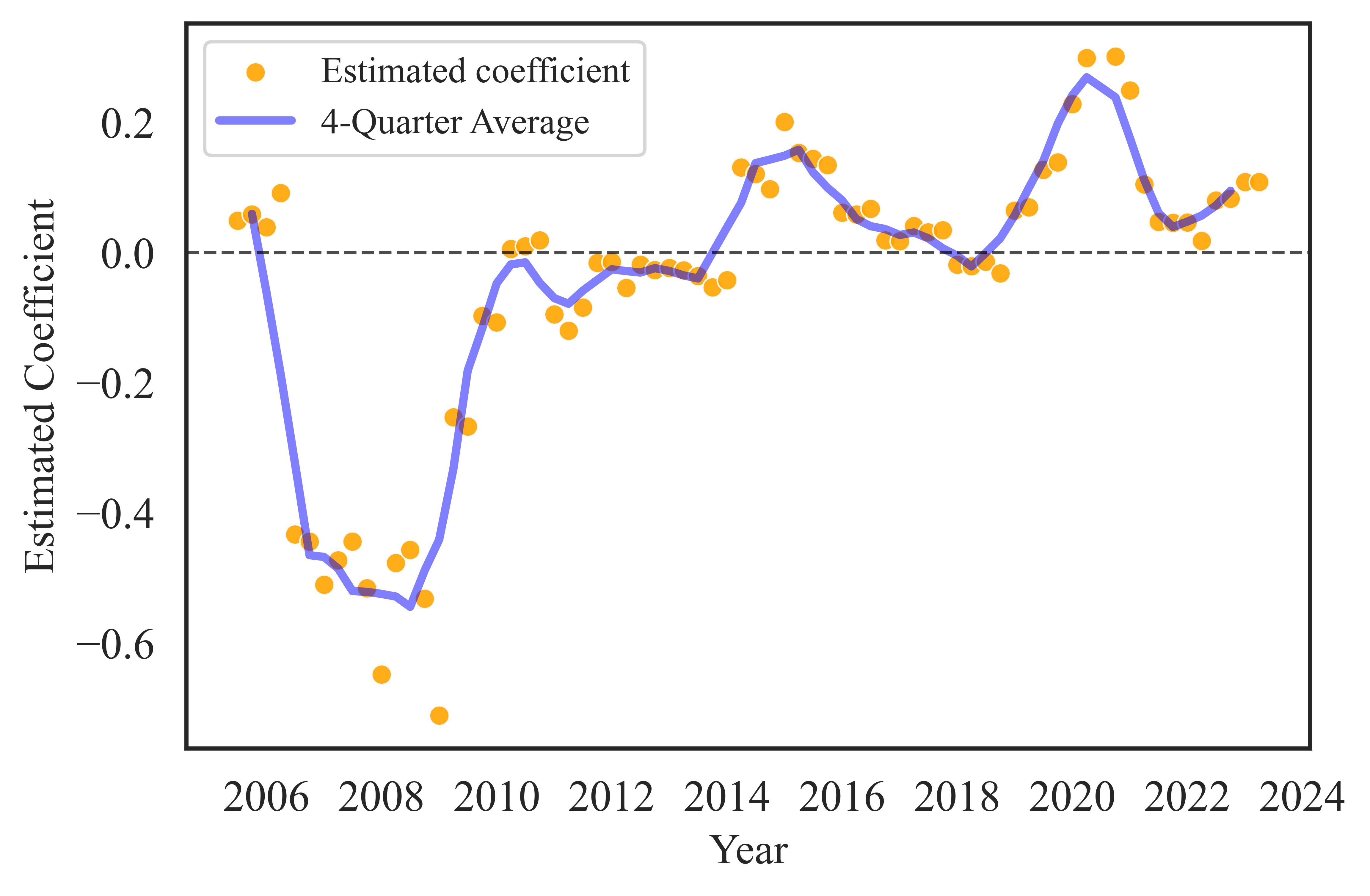}
    
    \justifying{
    \footnotesize{{Notes: i) The figure plots the coefficient $\beta_2$ for the regression $\pi_t = \beta_0 + \beta_1 \pi_{t-1} + \beta_2 \Delta y_t$ where $\pi$ is inflation and $\Delta y_t$ is the output gap.
    ii) The equation is estimated using an expanding window initialised at 20 quarters.}}}
    \label{fig: estimated_coef}
\end{figure}

\begin{figure}[ht!]
    \centering
    \caption{Inflation and Economic Slack: Structural shifts}
    \includegraphics[width=0.999\linewidth]{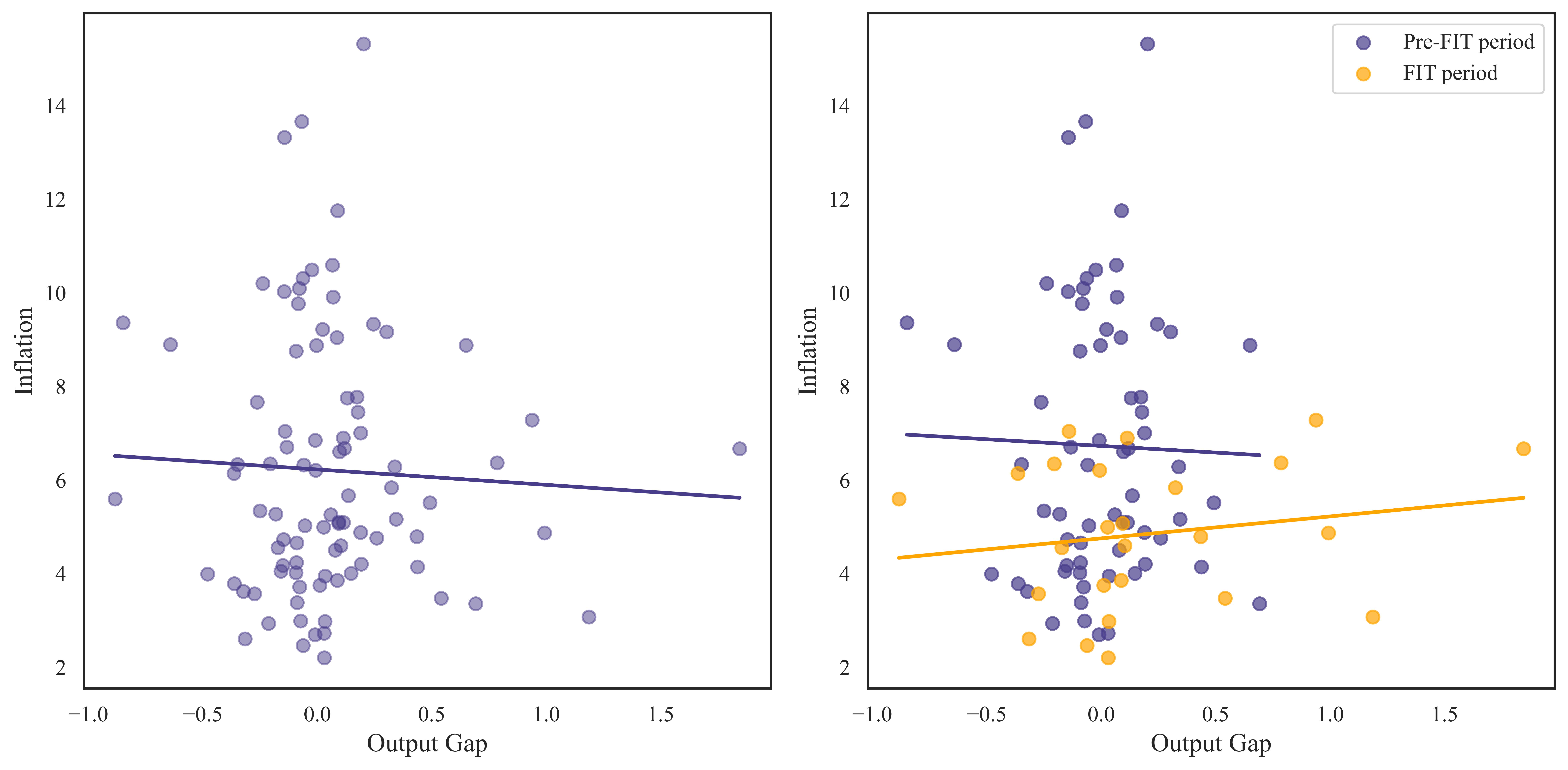}
   
   \justifying{
   \footnotesize{{Notes: i) The left panel shows the relationship between inflation and output gap for the entire sample. 
    ii) The right panel depicts how the relationship has altered following adoption of the Flexible Inflation Targeting (FIT) Regime since October 2016.
    iii) The charts exclude a negative outlier in output gap observed in Q2:2020 due to the Covid-19 pandemic and associated lockdowns.}}}
    \label{fig: scatt}
\end{figure}

Notwithstanding the success of monetary policy in taming inflation in India, forecasting inflation in India still remains a challenge for policymakers owing to the issues highlighted above. For instance, chart \ref{fig: estimated_coef} depicts the dynamic nature of relationship between output gap and inflation in India over the last two decades. While economic theory of the Phillips Curve suggests a positive relationship, the actual relationship in this case has oscillated over time, with the mean estimate being 0.11 across the whole 2000Q1 - 2023Q1 sample. Furthermore, Chart \ref{fig: scatt} highlights that the relationship between inflation and output gap in India may have undergone a structural break amidst the policy transition. The above evidence suggests that the PC relationship in the Indian context is complex which may not be adequately captured by simplistic, linear regression models that are conventionally used for modelling and forecasting purposes. Given these complexities, a more nuanced approach to inflation forecasting becomes crucial for policymakers to formulate effective monetary policy responses and ensure macroeconomic stability.

Therefore, in this paper, we use quarterly data ranging from Q1:2000 to Q1:2023, to train a host of linear, Tree-based and Neural Network-based models across various specifications of the Phillips curve. We then undertake a pseudo out-of-sample forecasting exercise to demonstrate that Tree-based models achieve significantly lower forecast errors compared to traditional linear regression models, including regularized models such as the LASSO model. This is reported across one- and multi-step ahead forecast horizons and different Phillips curve specifications. Tree-based models deliver superior predictive performance, even surpassing state-of-the-art Neural Network models. While neural networks offer advanced capabilities, their effectiveness is constrained by the requirement for large datasets, making them less suitable for contexts where data is limited. Our findings remain robust to several robustness checks.


Thereafter, we apply several \emph{explainable} ML techniques to a hybrid Phillips curve model (which includes both backward- and forward-looking components) to explore the nonlinear relationship between inflation and its structural determinants in the Indian case. To address the "black box" criticism of ML models, we use methods such as feature importance (FI), partial dependence plots (PDP), Shapley values and Shapley regressions \citep{shapley1953value, breiman2001random, friedman2001greedy, lundberg2017unified, strumbelj2010efficient, joseph2019parametric, buckmann2023interpretable}. 

In particular, we compare the Random Forest (RF) model with a Linear Regression (LR) benchmark, focusing on key aspects of the NKPC model, namely inflation expectations, lagged inflation, and the output gap. Our analysis suggests that inflation and its determinants, seen within the Phillips curve framework, share a highly nonlinear relationship. In terms of variable (feature) importance, both models highlight inflation expectations and lagged inflation as key drivers, but the RF model emphasizes expected inflation, while the LR model favours lagged inflation. The output gap, though absent from the LR model's top features, is captured by the RF model, suggesting it better identifies the nonlinear relationship between inflation and economic slack that is missed by the LR model. 

PDPs support these findings where inflation expectations display a non-linear, step-like pattern indicating thresholds; lagged inflation appears linear; and the output gap shows a state dependence based on the whether the output gap is positive or negative. We find that there exists more than one threshold -- in the range of 4.75 - 7.25 percent -- around which the level of inflation expectations has an asymmetric impact on realized inflation. Similarly, the influence of backward-looking component in the Phillips curve – lagged inflation – also increases sharply when the level of inflation crosses the range of 5.0 - 6.0 per cent. Two-way PDPs reveal that the RF model is able to capture interaction effect among these variables which contributes to its higher accuracy over the LR model. Furthermore, using a Shapley values-based approach, we provide a global interpretation of our model, finding an asymmetric response of headline inflation to high vs. low values of input features or variables. Shapley dependence plots highlight the non-linear relationships captured by the RF model in contrast to the linear LR model. Shapley regressions for the RF model confirm that expected inflation, lagged inflation, rainfall deviation, and the output gap are the primary drivers of inflation in India.  

In what follows, a brief survey of related literature on the Phillips curve as well as ML interpretation and inference is presented in Section \ref{sec:literature}. Our approach to model training, forecasting along with a brief discussion of explainable ML techniques is presented in Section \ref{sec:data}. The key findings of our paper including the forecasting analysis and model explanation using ML methods are discussed in Section \ref{sec:result}. Robustness analysis is presented in Section \ref{sec:robustness}. The paper concludes in Section \ref{sec:conclusion} while drawing contours for further research. 

\section{Related Literature} \label{sec:literature}

This study builds upon two primary strands of literature: one on the Phillips Curve, encompassing both theoretical and empirical research on inflation dynamics, and second on the application of machine learning (ML) and explainable AI (xAI) techniques for macroeconomic forecasting. By synthesizing these perspectives, we contribute to a more nuanced understanding of inflation dynamics particularly in the Indian context.

The Phillips Curve, which characterizes the trade-off between inflation and unemployment, was first empirically documented for the United Kingdom by \citet{phillips1958relation} and later extended to the US by \citet{samuelson1960analytical}. The introduction of inflation expectations into Phillips Curve model by \cite{phelps1967phillips} and \cite{friedman1968role} laid the foundation for the expectations-augmented Phillips Curve. This in turn led to the development of the micro-founded New Keynesian Phillips Curve (NKPC) with seminal contributions by \citet{taylor1980aggregate}, \citet{calvo1983staggered} and \citet{gali1999inflation}. Subsequent empirical studies, such as those by \citet{roberts1995new}, \citet{fuhrer1995phillips} and \citet{sbordone2002prices} tested and refined this framework. Empirical validity of the Phillips Curve, however, has been debated. Influential studies such as \citet{ball2011inflation} and \citet{stock2020slack} have questioned its robustness, arguing that structural changes in the economy—such as globalization, labor market rigidities, and higher degree of inflation anchoring—have weakened the relationship between inflation and economic slack. More recent studies, such as \citet{hazell2022slope} and \citet{ball2019phillips}, have sought to revive the Phillips Curve by incorporating nonlinearities and state-dependent effects in its modelling and estimation. More recently, \citet{doser2023inflation}, examine nonlinearities in the Phillips Curve using a threshold regression model and find that once inflation expectations—especially those of consumers—are properly accounted for, a linear specification cannot be rejected. The study shows that perceived nonlinearity often arises from mismeasured expectations rather than inherent structural breaks. While some episodes, like the “missing disinflation,” exhibit mild nonlinear effects, the model’s predictive power remains largely unchanged for key historical inflationary periods. Their findings underscore the central role of inflation expectations in shaping the Phillips Curve, challenging the necessity of nonlinear specifications in most economic conditions.

In the Indian context, early studies on the Phillips curve relationship encompass \cite{kapur2000price}, \cite{srinivasan2006modelling}, \cite{paul2009phillips}, \cite{mazumder2011stability}, \cite{patra2012monetary}, and \cite{kapur2013phillips}. Using various measures of prices and economic slack, most of these studies showed the presence of a distinct Phillips Curve in Indian data. Similarly, recent studies such as \cite{patra2014post}, \cite{ball2016understanding}, \cite{chinoy2016responsible}, \cite{behera2018phillips}, \cite{pattanaik2020inflation} and \cite{jose2021alternative} provide revised estimates for different Phillips curve specifications in the Indian context. These studies, in general, have also highlighted the role of supply-side shocks, especially those affecting food and fuel prices, in impacting inflation in India. \citet{mohanty2015determinants} and \citet{dua2021determinants} are related studies that offer a detailed analysis of inflation determinants for India. More recent studies, such as \citet{patra2021phillips}, explored whether the Phillips Curve remains relevant in India, particularly in the post-pandemic period. Their findings indicate that while the Phillips Curve relationship persists, its slope varies with the level of output gap, flattening when the gap is negative or low and steepening when it is high. A significant limitation in the Indian Phillips Curve literature is its heavy reliance on linear econometric models, which may fail to capture complex, nonlinear relationships between inflation and its determinants. Our study addresses this gap by leveraging machine learning (ML) methods, which allow for greater flexibility in modeling nonlinearities and structural breaks in the Phillips Curve relationship.

Recent advances in machine learning have led to the development of model-agnostic techniques for interpreting complex models and forecasts, enhancing their applicability in macroeconomic research \citep{guidotti2018survey, vilone2020explainable, molnar2020interpretable}). Methods such as Shapley values for local and global interpretation (\citealp{shapley1953value, strumbelj2010efficient, lundberg2017unified}; \citealp{lundberg2019consistent}), Partial Dependence Plots (\citealp{friedman2001greedy}), and Individual Conditional Expectations (\citealp{goldstein2015peeking}) are increasingly used to improve the explainability of machine learning models. In macroeconomics, ML-based forecasting has gained serious traction, especially in case of inflation, GDP and unemployment forecasting \citep{nakamura2005inflation, chakraborty2017machine, hall2018machine, almosova2023nonlinear, malladi2023benchmark}. Recent advancements in macroeconomic forecasting have emphasized the need for models capable of handling the complexities inherent in macroeconomic data, including non-linearities, long-range dependencies, and structural breaks. 

Several recent studies have also applied these methods to forecast inflation in India \citep{pratap2019macroeconomic, singh2022inflation, sengupta2024forecasting}. In particular, \citet{sengupta2024forecasting} introduce a novel algorithm, namely the Filtered Ensemble Wavelet Neural Network (FEWNet), specifically designed to address such challenges. Their model decomposes inflation data into high- and low-frequency components using wavelet transforms while integrating additional exogenous factors, such as economic policy uncertainty and geopolitical risk to improve forecast accuracy. By leveraging wavelet-transformed inputs and filtered variables within an ensemble of autoregressive neural networks, FEWNet effectively captures intricate data patterns and enhances its adaptability to macroeconomic forecasting tasks. The study, applied to the BRIC nations, demonstrated that FEWNet consistently outperforms both traditional linear statistical models and state-of-the-art non-linear algorithms in producing robust long-term (24-month-ahead) inflation forecasts. 

Beyond macroeconomic forecasting, ML and xAI techniques are also being used broadly in economic and financial domains to derive fresh policy insights \citep{sun2024deep}. Using artificial neural networks and SHapley Additive exPlanations, they uncover a nonlinear U-shaped relationship between the digital economy and energy productivity, highlighting regional heterogeneity, lagging effects, and spatial dependencies. Their findings demonstrate how ML-driven insights can enhance policy decisions by capturing dynamic interactions often overlooked in traditional econometric models. Additionally, research on green housing policies offers insights into how economic incentives and policy interventions shape behavioral responses, which can be analogously applied to monetary policy and inflation dynamics. Similarly, \citet{li2023social} develop a four-party evolutionary game model to examine interactions between government, realtors, and residents under different policy scenarios. Their findings demonstrate that government incentives and regulatory measures influence private sector decisions, leading to market-driven sustainability outcomes. This aligns with our study’s emphasis on state-market interactions in inflation dynamics, where monetary policy interventions affect inflation expectations and labor market responses, reinforcing the importance of nonlinear modeling approaches. 

This study advances the Phillips Curve literature in India by addressing key limitations of traditional linear models, which often fail to capture structural changes, supply shocks, and thresold effects in the data. While some studies incorporate state-dependent effects, they lack methodological rigor in modelling nonlinearity and time variation. By integrating ML techniques within the New Keynesian Phillips Curve (NKPC) framework, we provide a more flexible and theory-consistent approach to forecasting inflation. Our study further enhances interpretability by employing explainable AI methods such as Shapley values, Partial Dependence Plots, and Individual Conditional Expectations, offering policymakers with deeper insights into the key drivers of inflation and their nonlinear interactions. This synthesis of Phillips Curve theory, ML-based forecasting, and explainable AI strengthens our understanding of inflation dynamics in emerging economies, ensuring both improved predictive accuracy and greater policy relevance.

\section{Empirical Strategy and Data} \label{sec:data}
The empirical analysis is divided into three parts. In the first step, we train (estimate) a mix of linear regression and non-linear ML models on various specifications of the Phillips curve, namely a backward-looking PC, a purely forward-looking PC and a hybrid PC which incorporates both backward- and forward-looking components. In the second step, we generate multi-step ahead forecasts of inflation using each of the model-specification combination on a test (out-of-sample) dataset. We compare the forecast accuracy of these models against a simple Random Walk (RW) benchmark. The last part focuses on applying various explainable ML techniques to explain the predictions generated by our best performing model. More importantly, this part focuses on analyzing the non-linear relationship between inflation and its determinants in the Indian context. 

\subsection{Model Training and Testing} 
Following the literature on Phillips curve, three specifications are trained and tested – an adaptive PC, a purely forward-looking PC and a hybrid PC as shown below in equations (1), (2) and (3), respectively:

\begin{equation}
\pi_t = \beta_0 + \beta_1\pi_{t-1} + \sum_{k=1}^{k=4} \gamma_{t-k}y_{t-k} + \sum_{k=0}^{k=4} \delta_{t-k}X_{t-k} + \varepsilon_t 
\end{equation}

\begin{equation}
\pi_t = \beta_0 + \beta_1E_t\pi_{t+1} + \sum_{k=1}^{k=4} \gamma_{t-k}y_{t-k} + \sum_{k=0}^{k=4} \delta_{t-k}X_{t-k} + \varepsilon_t
\end{equation}

\begin{equation}
\pi_t = \beta_0 + \beta_1\pi_{t-1} + \beta_2E_t\pi_{t+1} + \sum_{k=1}^{k=4} \gamma_{t-0}y_{t-k} + \sum_{k=1}^{k=4} \delta_{t-k}X_{t-k} + \varepsilon_t 
\end{equation}

where $\pi_t$ is inflation, $E_t \pi_{t+1}$ is the inflation expected in $t+1$ period at time $t$, $y$ is the output gap and $X$ is a vector of controls variables, which mainly consists of supply-related indicators. Following the literature on India, X consists of changes in exchange rate, crude oil and rainfall deviation in the present case. The adaptive PC assumes the data generating process of inflation to be backward-looking i.e., determined by its own lagged values. Similarly, the forward-looking PC models inflation to be primarily driven by expected level of future inflation. Finally, the hybrid specification of the Phillips curve considers the inflationary process to contain both backward- and forward-looking components. 

The Phillips curve, shown in its linear form in equations (1) - (3), can be described in a nonlinear functional form as shown below. In this case, the nonlinear function \(F(.)\) is approximated by a machine learning model. 

\begin{equation}
    \pi_t = \left\{ F\left(\pi_{t-1}, y_{t-k}, \mathbf{X}_{t-k}\right) \right\}
\end{equation}

We employ a diverse set of models for inflation forecasting, categorised into linear models, non-linear models, and neural network-based models. Linear models include Linear Regression, Ridge Regression, and LASSO, each capturing relationships between predictors and inflation. Ridge Regression adds an L2 penalty to shrink the coefficients, thereby reducing variance, while LASSO uses an L1 penalty to perform both regularisation and feature selection by setting some coefficients to zero. Non-linear models, such as Random Forest and XGBoost, improve upon decision trees by either averaging predictions from bootstrapped trees (Random Forest) or building trees sequentially with gradient boosting (XGBoost). These models can capture more complex, non-linear patterns in the data. Neural network-based models include N-BEATS, which uses a fully connected architecture to decompose forecasts into trend and seasonality components, and N-HITS, which extends N-BEATS by incorporating hierarchical interpolation filters. We also use BlockRNN with LSTM cells, which are effective at capturing long-term dependencies in sequential data. These models can also incorporate external covariates, enhancing their ability to account for exogenous factors influencing the forecast. Additional details regarding these models are presented in the Appendix.

In practical macroeconomic forecasting, we are usually concerned with next period forecasts while using information available up to the current period. To recreate this scenario, we conduct an iterative forecasting horse-race between competing models. This is achieved by fitting a model on data up to quarter $q$ and then predicting the target variable for $q+1^{th}$ quarter. In the next step, the model is fitted on data up to quarter $q+1$ and prediction for $q+2^{nd}$ quarter is made and so on until we extinguish the test set. This process is repeated for each model. To make this exercise as close to real time forecasting as possible, output gap and expected inflation are also calculated iteratively by employing information only up to last available quarter in each iteration. The pseudo-algorithm is presented as below. In the present case, our sample consists of 93 quarters out of which the test set comprise of the last 24 quarters (6 years).

\begin{algorithm}
\caption{Forecasting HorseRace}
\vspace{-0.01cm}
\small
\label{Algorithm_1}
\begin{algorithmic}
\Function{1QuarterAheadForecasts}{$D, Q, M$}
\State \textbf{Input:} Full dataset $D$, total quarters $Q$, set of models $M$
\State \textbf{Output:} Set of one-period-ahead forecasts $F$
\State Initialize $q \gets$ initial quarter for first forecast
\State Initialize $F \gets \emptyset$
\While{$q < Q-1$}
    \State Extract subset $D_q \gets D[1:q]$
    \State $\text{OutputGap} \gets \text{OutputGap.UCM}(D_q)$
    \State $\text{TrendInflation} \gets \text{TrendInflation.UCM}(D_q)$
    \State Augment $D_q$ with $\text{OutputGap}$ and $\text{TrendInflation}$
    \State $\text{ExpectedInflation} \gets \text{Lead}(\text{TrendInflation}, 1)$
    \ForAll{$m \in M$}
        \State Train model $m$ on augmented $D_q$
        \State $f_m \gets \text{Forecast}(m, \text{horizon}=1)$
        \State $F \gets F \cup \{f_m\}$
    \EndFor
    \State $q \gets q + 1$
\EndWhile
\State \Return $F$
\EndFunction
\end{algorithmic}
\end{algorithm}

Finally, we compute several error metrics namely root mean squared error (RMSE), median absolute relative error (MdRAE), Symmetric Mean Absolute Percentage error (SMAPE)  and Theil's U2 statistic for each of the competing models for comparison. Computational details for each of these metrics is provided in Appendix A. 

\subsection{Model Explanation}

Interpretability is often a big hurdle for using a machine learning (ML) model in economics. Classical econometric models usually assume a data-generating process beforehand, which a priori imposes constraints on the model but allow for easy interpretation of the model, including its coefficients and forecasts. On the other hand, ML algorithms usually employ nonlinear and non-parametric approaches which results in poor interpretability of the model. Understandably, by choosing nonlinearity over simpler linear methods, ML algorithms tend to be complex and not-so straightforward to understand. This section briefly describes the model-agnostic, interpretability techniques used in the paper.

\subsubsection{Permutation Importance}

Permutation importance is a model-agnostic method used to estimate the importance of individual features in predictive models by measuring the decrease in performance when a feature's values are randomly shuffled. Given a trained model \( f \) and a performance metric \( M \), the baseline performance \( M(f, X, y) \) is first computed on the original dataset \( X \). For each feature \( X_j \), its values are permuted(i.e. randomly shuffled), generating a new dataset \( X^{\text{perm}}_j \), while all other features remain unchanged. The model performance \( M(f, X^{\text{perm}}_j, y) \) is then evaluated on this permuted dataset. The importance of feature \( X_j \), denoted as \( I(X_j) \), is defined as the difference in model performance before and after permutation:

\begin{equation}
    I(X_j) = M(f, X, y) - M(f, X^{\text{perm}}_j, y)
\end{equation}

This method quantifies the contribution of each feature to the model's predictions. Features with larger importance values have a more significant impact on the model, as their randomization leads to a greater reduction in performance. Permutation importance is widely applicable and does not require retraining the model, making it computationally efficient and suitable for complex models, such as random forests or neural networks. However, it can be sensitive to feature interactions and correlations, which may result in an underestimation of a feature's true importance.

\subsubsection{Partial Dependence Plots (PDP) and Individual Conditional Expectation (ICE) plots}

Partial Dependence Plots (PDP) and Individual Conditional Expectations (ICE) are model-agnostic techniques used to visualize the effect of one or more features on the predicted outcome of a machine learning model. PDPs help estimate how a feature influences the model's predictions by averaging the effect of that feature over all possible values of other features. This process marginalizes the effects of other features, providing an expected outcome for any given value of the feature of interest.

Mathematically, given a trained model \( f \), feature set \( X \), and target variable \( y \), the partial dependence of a feature \( X_j \) on the model prediction is computed as the expected prediction of the model, averaging over all other features \( X_{-j} \) (i.e., all features except \( X_j \)). This expectation is taken with respect to the marginal distribution of \( X_{-j} \), denoted as \( P(X_{-j}) \). The partial dependence function is defined as:

\begin{equation}
    PD(X_j) = \mathbb{E}_{X_{-j} \sim P(X_{-j})} [f(X_j, X_{-j})]
\end{equation}

where \( \mathbb{E}_{X_{-j} \sim P(X_{-j})} \) denotes the expectation over the distribution of the remaining features. In essence, the PDP reflects the average predicted value for different values of \( X_j \), accounting for the variability of all other features.

In contrast, ICE plots provide a more granular view by showing the individual predictions for each data point in the dataset. Instead of averaging over the distribution of other features, ICE plots keep all other features fixed and vary only \( X_j \) for each observation, offering insights into how different instances react to changes in \( X_j \).

The calculation of Partial Dependence Plots (PDPs) assumes that the features are uncorrelated, an assumption that may not always hold in practice. Consequently, when a feature is highly correlated with other features, the computation of its PDP often involves averaging predictions over synthetic data points that may be implausible or rare in real-world scenarios. This can introduce significant bias in the estimation of the feature’s true effect, potentially leading to misleading interpretations.

When considering interactions between two features \( X_j \) and \( X_k \), a 2-way PDP can be employed. This plot visualizes the joint effect of two features on the model prediction by calculating the expected predictions over a grid of values for \( X_j \) and \( X_k \), while marginalizing over all remaining features \( X_{-j,k} \):

\begin{equation}
    PD(X_j, X_k) = \mathbb{E}_{X_{-j,k} \sim P(X_{-j,k})} [f(X_j, X_k, X_{-j,k})]
\end{equation}

This method helps reveal whether and how the interaction between the two features influences the model's predictions, which cannot be captured by single-feature PDPs.

In summary, PDPs provide a global view of feature effects averaged over the distribution of other features, while ICE plots reveal individual variations. The combination of both techniques helps in understanding both global trends and instance-specific behaviour in complex machine learning models.

\subsubsection{Shapley Values}

Originally introduced in cooperative game theory, Shapley values provide a method to allocate a joint payoff obtained in a cooperative game to the individual players of a coalition based on their contributions \citep{shapley1953value}. The concept of Shapley values was introduced to machine learning by \cite{strumbelj2010efficient} as a way of attributing predictions in a supervised model to its constituent features. In the context of ML models, shapley values $\phi_k(x_i;f)$ for a given feature $k$ and prediction $x_i$ measure the marginal contribution of feature $k$ to the prediction, averaged over all possible subsets of the features. Formally, for a model $f$, the Shapley value for feature $k$ in prediction $x_i$ is given by:

\begin{equation}
    \phi_k(x_i;f) = \sum_{S \subseteq N \setminus \{k\}} \frac{|S|!(n - |S| - 1)!}{n!} \left[ f(x_i|S \cup \{k\}) - f(x_i|S) \right]
\end{equation}

where $N \setminus \{k\}$ represents all subsets of features excluding $k$, $S$ denotes the number of variables included in that subset and $n$ is the total number of features.

\textbf{Intuitive Description:} The formula for Shapley values can be understood by imagining each feature as a player in a cooperative game. The goal is to fairly distribute the payout (the model's prediction) among all the features based on their contributions. For any feature $k$, we calculate how much adding that feature changes the model's prediction across all possible subsets $S$ of the other features. This change is weighted by the number of subsets in which feature $k$ could potentially appear, ensuring a fair contribution is assigned to each feature, regardless of the order in which they are considered. The Shapley value $\phi_k(x)$ is essentially the weighted average of these marginal contributions across all subsets.

Shapley regression, introduced by \cite{joseph2019parametric}, is a statistical inference framework that leverages Shapley values to evaluate the importance of features in machine learning models. The approach allows for conventional parametric inference in nonlinear models by performing a regression of the outcome variable on the Shapley values of the features. This makes it possible to test the statitical significance of individual features in machine learning predictions often treated as "black boxes". The regression model is formulated as:

\begin{equation}
    y_i = \beta_0 + \sum_{k=1}^n \beta_k \phi_k(x_i) + \epsilon_i
\end{equation}

where \( y_i \) represents the observed outcome for the \( i \)-th observation, while \( \beta_0 \) denotes the intercept of the model. The term \( \phi_k(x_i) \) refers to the Shapley value for the \( k \)-th feature in the \( i \)-th observation, capturing the contribution of feature \( k \) to the model's prediction. The corresponding regression coefficient, \( \beta_k \), quantifies the strength of the relationship between the Shapley value and the outcome.

This regression allows us to estimate the significance of each feature by testing the hypothesis:

\begin{equation}
    H_0: \beta_k \leq 0 | x_i \in \Omega
    \label{h0Shapleyregression}
\end{equation}
where  $\Omega \in \mathbb{R}^n$ is the model input space, where \( n \) represents the number of input features.

The key idea of Shapley regression is to capture the importance of each feature in a nonlinear ML model in a way that is analogous to the coefficients in a linear regression model. The Shapley value \( \phi_k(x_i) \) quantifies the contribution of feature \( k \) to the prediction for the \( i \)-th observation. By regressing the observed outcome on these contributions, we are able to make inferences about the importance of features as if we were working with a linear model. The advantage of this framework, as shown by \cite{joseph2019parametric}, is that it allows for parametric statistical inference on machine learning models, which is typically difficult due to their nonlinearity and complexity.

Additionally, Shapley regression provides a mechanism for calculating the Shapley share coefficients $\Gamma_k$, which measure the relative contribution of each feature across the entire dataset. The Shapley share for feature \( k \) is defined as:

\begin{equation}
    \Gamma_k = sign \frac{|\phi_k(x_i)|}{\sum_{j=1}^n |\phi_j(x_i)|}^{(*)}
\end{equation}

where $sign$ is the sign of coefficients when regressing $y$ on $x$ and (*) indicate the confidence level  with which we can reject $H_0$ in equation (\ref{h0Shapleyregression}). The interpretation of $\Gamma_k$ is also similar to that of regression coefficient, as it measures the strength, direction and confidence in alignment with the target variable. However, it should be noted that $\Gamma_k$ cannot be interpreted as the marginal effect, unless the model is linear.

Shapley values possess desirable properties such as consistency and efficiency, making them a powerful tool for feature attribution in machine learning models (\citealp{lundberg2017unified}). Furthermore, Shapley regressions bridge the gap between predictive modeling and statistical inference, providing insights into feature importance and significance in a rigorous manner.

\subsection{Data}
The data employed in the study begins in 2000Q1 and ends in 2023Q1. Table \ref{tab:data-desc} presents a brief description of the data.


\begin{table}[h]
\centering
\caption{Data Description}
\scalebox{0.9}{
\begin{tabular}{p{2.5cm}|p{8cm}|p{2.5cm}}
\hline
\textbf{Variable} & \textbf{Description} & \textbf{Source} \\
\hline
Inflation & Quarterly inflation is calculated as the year-on-year (y-o-y) change in the average value of the Consumer Price Index(CPI): Combined in a quarter:  
$\text{Inflation}_t = \frac{\text{CPI}_t - \text{CPI}_{t-4}}{\text{CPI}_{t-4}} \times 100$.  
The quarterly series is generated by combining the CPI-Industrial Workers series (Base: 2001) with the current CPI-Combined series (Base: 2012). & RBI-DBIE \\
\hline
GDP & Real GDP series are constructed by linking series across base years 1999-2000, 2004-05, and 2011-12. & RBI-DBIE \\
\hline
Rainfall deviation & Deviation of rainfall from its long-period average (LPA) is used to measure monsoon performance. & IMD \\
\hline
Crude prices & Equally weighted average of Brent, WTI, and Dubai spot prices expressed in year-on-year growth terms. & CEIC \\
\hline
USD/INR rate & Year-on-year growth of quarterly average values of the USD/INR exchange rate. & CEIC \\
\hline
\multicolumn{3}{p{0.9\textwidth}}{\footnotesize{\textbf{Note:} RBI-DBIE - Reserve Bank of India-Database on Indian Economy; IMD - Indian Meteorological Department.}}
\end{tabular}}
\\\vspace{1ex}
\raggedright
\label{tab:data-desc}
\end{table}

We compute the output gap using an Unobserved Component Model (UCM) on seasonally adjusted quarterly GDP series\footnote{Seasonal adjustment was made using the X-13 ARIMA model.}. Similarly, trend inflation is calculated by fitting a UCM model to the quarterly inflation series.\footnote{See \cite{watson1986univariate} and \cite{harvey1993detrending} for more details on the UCM model.} The trend component of inflation – shifted forward one-time step – is used as a measure of inflation expectations. 

\begin{figure}[ht]
    \centering
     \caption{Inflation and its determinants}
     \label{chart:economic-indicators1}
    \includegraphics[scale=0.66]{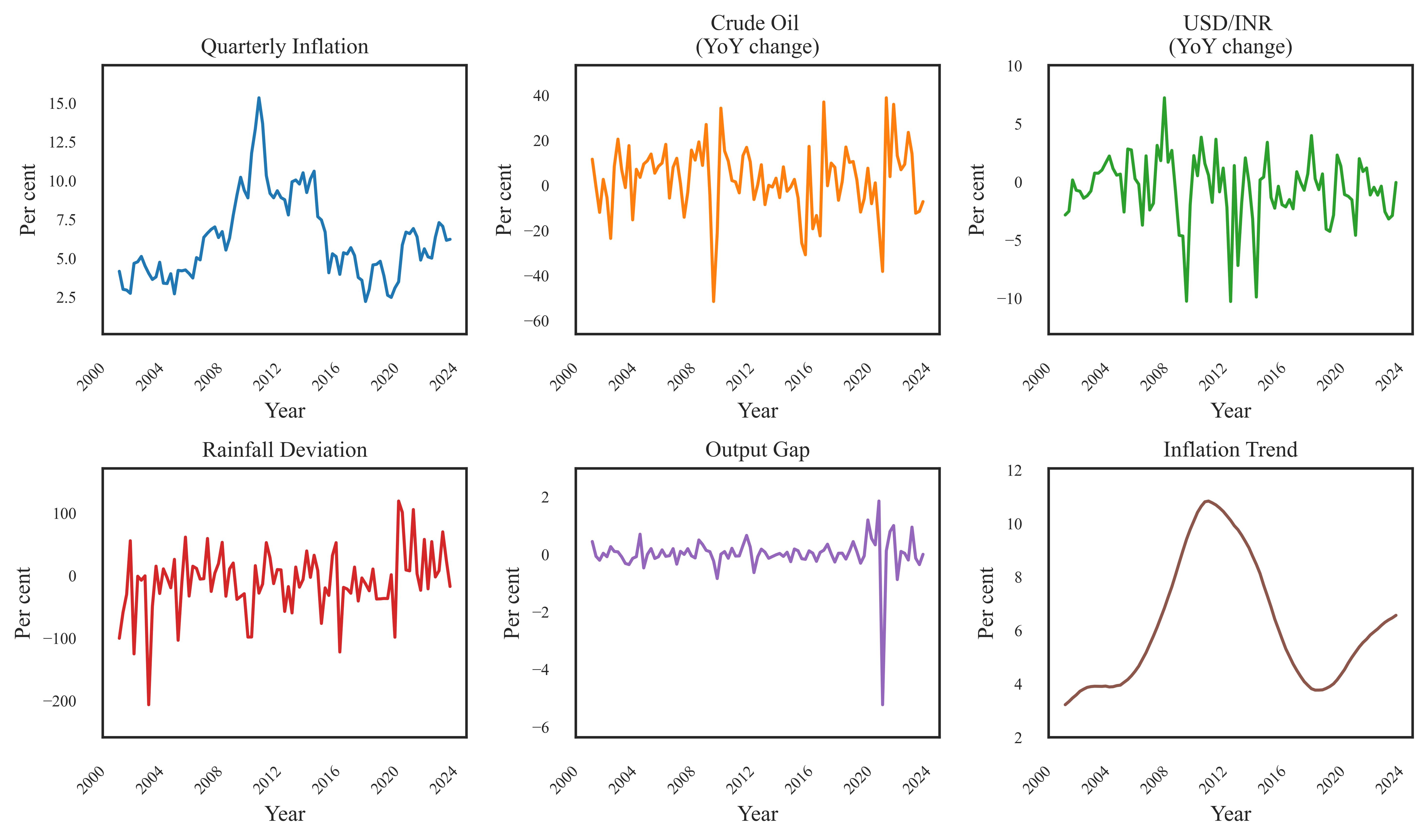}
    \captionsetup{labelformat=simple, labelsep=colon}
   
   \justifying{
   \footnotesize{{Note: Trend inflation and output gap is estimated using an Unobserved Components Models on logarithm of GDP and Inflation respectively.}}}
\end{figure}

Given that inflation expectations are unobserved, there is no consensus in literature about the best measure of inflation expectations. Market based methods, derive inflation expectations using inflation-indexed bonds, inflation swaps and inflation options. However, such measures may be noisy and reflect risk premia. Further, there doesn’t exist a liquid market for such instruments in the context of India. Secondly, survey measures may also be employed to derive inflation expectations. \cite{cecchetti2007understanding} provided evidence that survey inflation forecasts are correlated with future trend inflation, measured using a \cite{stock2007why} UCSV model. However, in the context of India, a long, continuous time-series of survey-based data on inflation expectations is not available or is available at an uneven frequency. Further, several studies have highlighted the need for debiasing inflation expectations derived from responses in the Survey of Households data (\citealp{muduli2022assessing}). Amid these data issues, researchers are forced to look at “second best” options. For example, \cite{patra2010inflation} fit an ARMA model to observed inflation and use the one-period ahead forecasts as a measure of inflation expectations. Some others like \cite{bicchal2019rationality} have used Google Trends data to derive inflation expectations for India. We select our measure of inflation expectations keeping in mind the issues presented above. Chart \ref{chart:economic-indicators1} plots the data series employed in the study.

The analysis of the statistical properties of key macroeconomic variables, summarized in Table 2, reveals several critical characteristics that provide insights into the underlying dynamics of the data and inform the selection of appropriate forecasting methodologies. The data exhibits significant variability, particularly in supply-side factors such as crude oil prices, rainfall deviations, and the output gap, as indicated by high coefficients of variation and entropy values. Most variables, including quarterly inflation, crude oil prices, USD/INR exchange rate, and rainfall deviations, demonstrate the presence of long-range dependence, as identified by the Hurst exponent. In particular, the output gap is the only variable that does not exhibit long-range dependence, reflecting its distinct structural characteristics within the data set.
Tests for non-linearity, including Tsay’s test (\citealp{tsay1986nonlinearity}) and Keenan’s one-degree test (\citealp{keenan1985tukey}), indicate that the majority of the variables, except for crude oil prices and trend inflation, exhibit non-linear dynamics. This non-linearity emphasizes the importance of utilizing flexible forecasting models capable of capturing these complex relationships. Additionally, all variables are confirmed to be trend-stationary based on the Kwiatkowski–Phillips–Schmidt–Shin (KPSS) test (\citealp{kwiatkowski1992testing}), ensuring their suitability for econometric modeling. The Ollech and Webel test (\citealp{ollech2023random}) identifies no seasonality in any of the variables, further simplifying the structural assumptions required for modeling. Skewness and kurtosis values reveal that variables such as inflation and output gap are positively skewed, reflecting asymmetric distributions, while crude oil prices exhibit negative skewness.
Outlier detection using the Bonferroni outlier test (\citealp{weisberg1982residuals}) based on Studentized residuals reveals the presence of outliers in the supply-side factors (crude oil and exchange rates) and the output gap, indicating the occurrence of sudden and significant shocks in these variables. These findings collectively highlight the complexity and heterogeneity of the dataset, necessitating advanced forecasting approaches, particularly those that can accommodate non-linearities, long-range dependence, and the presence of outliers. This comprehensive analysis serves as a foundation for developing robust econometric and machine learning-based forecasting models tailored to the intricate dynamics of these macroeconomic variables.

\begin{table}[h!]
\centering
\caption{Global Characteristics}\label{tab:summ_stats}
\scalebox{0.62}[0.7]{  
\begin{tabular}{l|r|r|r|r|r|r}
\hline
\textbf{Variables} & \textbf{Quarterly Inflation} & \textbf{Crude Oil (YoY)} & \textbf{USD/INR (YoY)} & \textbf{Rainfall Deviation} & \textbf{Output Gap} & \textbf{Trend Inflation} \\
\hline
Observations & 89 & 89 & 89 & 89 & 89 & 89 \\
\hline
Min. Value & 2.202 & -51.591 & -10.299 & -206.902 & -9.303 & 1.832 \\

Max. Value & 15.315 & 38.762 & 7.206 & 118.800 & 6.515 & 8.399 \\

Q1 & 4.175 & -5.716 & -1.977 & -29.422 & -2.232 & 3.487 \\

Median & 5.593 & 3.102 & -0.390 & -5.386 & -0.412 & 4.514 \\

Mean & 6.268 & 2.395 & -0.592 & -7.956 & -0.470 & 4.627 \\

Q3 & 7.752 & 10.688 & 1.167 & 14.558 & 1.257 & 5.628 \\

Entropy & 4.422 & 6.743 & 8.182 & 7.196 & 8.094 & 5.330 \\

CoV & 43.687 & 647.962 & -493.305 & -627.896 & -1302.579 & 27.875 \\
\hline
Skewness & 0.920 & -0.469 & -0.978 & -0.660 & -0.506 & 0.516 \\

Kurtosis & 0.514 & 1.210 & 2.361 & 2.435 & 2.474 & 0.541 \\

Non-Linear & Non-Linear & Linear & Non-Linear & Non-Linear & Non-Linear & Linear \\

LR Dependence & 0.764 & 0.576 & 0.567 & 0.606 & 0.491 & 0.829 \\

Seasonality & Non-Seasonal & Non-Seasonal & Non-Seasonal & Non-Seasonal & Non-Seasonal & Non-Seasonal \\

Stationarity & Trend - Stationary & Trend - Stationary & Trend - Stationary & Trend - Stationary & Trend - Stationary & Trend - Stationary \\

Outlier & No outlier detected & 1 outlier detected & 2 outliers detected & 1 outlier detected & 1 outlier detected & No outlier detected \\
\hline
\end{tabular}}
\end{table}

\section{Results} \label{sec:result}
\subsection{Forecasting Performance and Benchmark Comparison}
\begin{figure}[htbp!]
\caption{Forecast Error Comparison \textit{vis-à-vis} a Random Walk Baseline}
\begin{subfigure}
  \centering
  \includegraphics[width=1\linewidth]{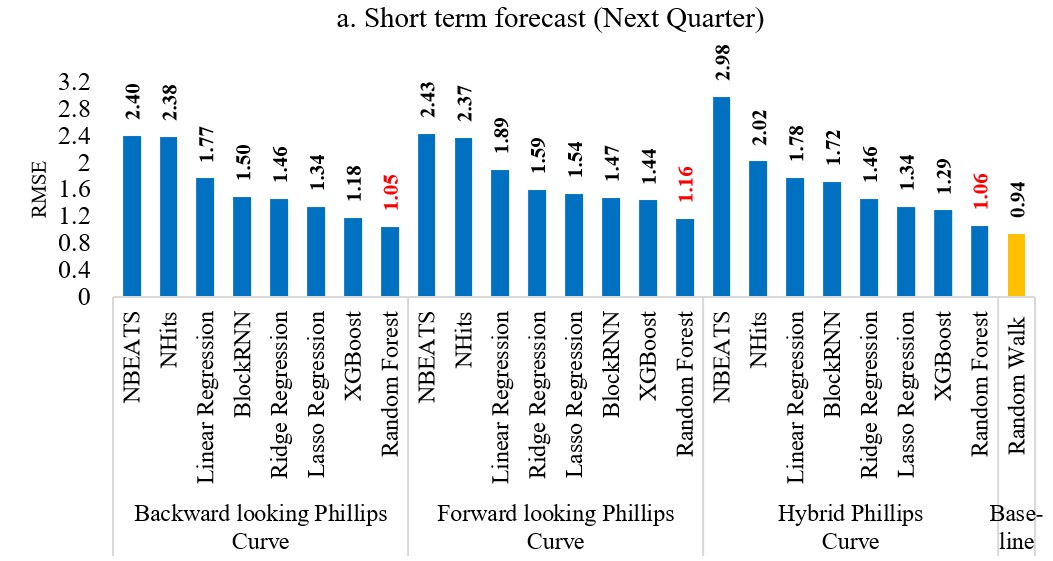}
  \label{fig:sfig1}
\end{subfigure}%
\begin{subfigure}
  \centering
  \includegraphics[width=1\linewidth]{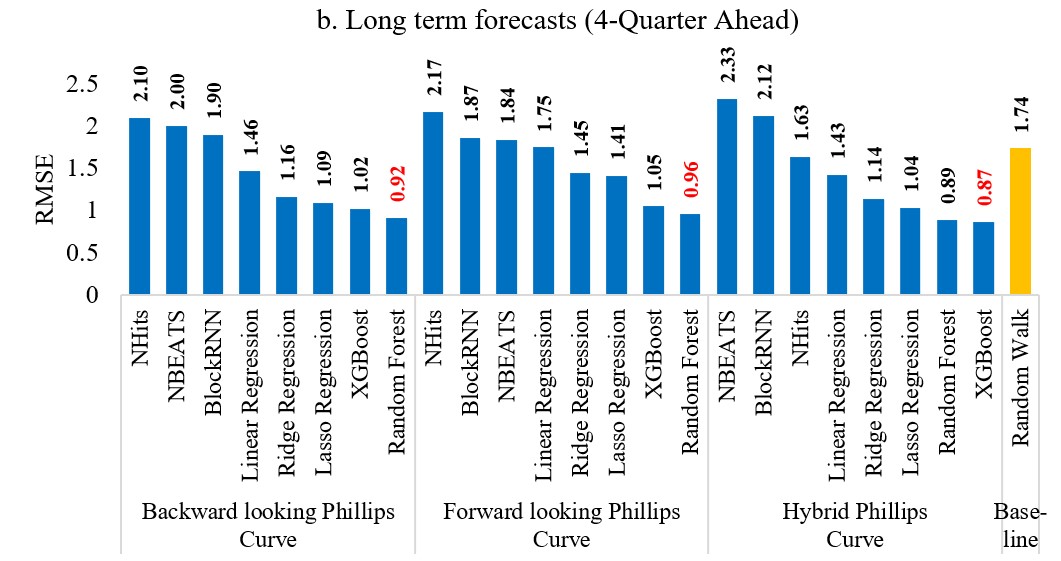}
  \label{fig:sfig2}
\end{subfigure}

\label{fig:fig_forecast_acc}
\footnotesize{{Notes: i) RMSE is calculated using forecast errors for 24 quarters for 1-quarter ahead and 21 quarters for 4-quarter ahead in the test set. Adaptive PC, Forward-looking NKPC and Hybrid NKPC refer to the specifications described in equations (1), (2) and (3), respectively. Shrinkage parameter for Ridge and LASSO are optimised through cross-validation at each iteration. Random Forest and XGBoost are optimised using five-fold cross validation with a Grid search algorithm.

ii) Best performing model’s RMSE is printed in red boldface.}}
\end{figure}

As described earlier, we undertake a pseudo real-time out-of-sample forecasting exercise and compute the forecast errors across different models and specifications. The root mean square error (RMSE) for the five models and three specifications for 1- and 4-quarter ahead forecast horizons\footnote{Details forecast error metrics are presented for 1 up to 4-quarter ahead forecasts in Appendix A: Table A1 through A4.} presented below in Chart \ref{fig:fig_forecast_acc}. Further, RMSE of a benchmark Random walk model is also presented for comparison. 

The results reveal that machine learning models, particularly Random Forest and XGBoost, consistently outperform both linear models (Linear Regression, Ridge Regression, and LASSO) and complex neural network models (NBEATS, NHits, and BlockRNN) in forecasting CPI inflation across short-term (1-quarter ahead) and longer-term (4-quarter ahead) horizons. Chart 4 provides a detailed comparison of root mean squared error (RMSE) values across different forecasting approaches for the three Phillips Curve specifications. For 1-quarter ahead forecasts, Random Forest achieves the lowest RMSE under the backward-looking Phillips Curve (1.05) and hybrid Phillips Curve (1.06), while XGBoost delivers competitive performance under the forward-looking Phillips Curve (1.16). The 4-quarter ahead results exhibit a similar pattern, with Random Forest achieving the lowest RMSE under the hybrid Phillips Curve (0.89) and backward-looking Phillips Curve (0.92), and XGBoost leading under the forward-looking Phillips Curve (0.96). These findings underscore the exceptional performance of Random Forest and XGBoost in capturing the complex, non-linear dynamics of inflation, surpassing both simpler linear models and more complex neural network-based approaches. Additionally, the Random Walk model, serving as a baseline, exhibits strong short-term accuracy but is less robust for multi-step forecasts, further highlighting the predictive power of the machine learning approaches.

The superiority of ensemble ML models suggests that their ability to capture non-linearities and complex interactions between variables provides a significant advantage for macroeconomic forecasting. This result highlights the potential for machine learning techniques to enhance the accuracy of macroeconomic models, particularly in contexts where relationships between variables may be more complex than what linear methods can adequately represent.

The Random Walk model, however, outperforms all other models for one-quarter ahead predictions with the lowest RMSE (0.94). Despite this, its performance deteriorates substantially for the 4-quarter ahead forecasts, where it has the highest RMSE (1.74). While the Random Walk model can provide accurate short-term predictions, in the current case, it neither offers any insights into the structural determinants of inflation nor help explain the factors driving inflation forecasts generated at any given time. This limitation makes it less useful for policy analysis where understanding the underlying drivers of inflation or unemployment is critical. This is true even for other univariate time-series models, such as ARIMA models. In contrast, models based on the Phillips curve framework not only provide forecasts but also offer interpretability, helping policymakers understand the economic forces at play, which is essential for informed decision-making.
Overall, Random Forest model outperforms in the one-quarter ahead forecasts while XGBoost stands as the best model for 4-quarter ahead forecasts. Specifically, Random Forest model achieves 20-40 percent improvement over the linear and regularised linear regression model. Overall, the hybrid NKPC estimated using the Random Forest algorithm provides the best forecasts for one-quarter ahead forecasts.

\subsection{Machine Learning Model Explanation}

Based on the forecasting results, we select the hybrid Phillips curve specification with the Random Forest (RF) model for further analysis and comparison with Linear Regression (LR). 

\begin{figure}[h!]
 \caption{Variable Importance}
    \centering
    \includegraphics[width=0.999\linewidth]{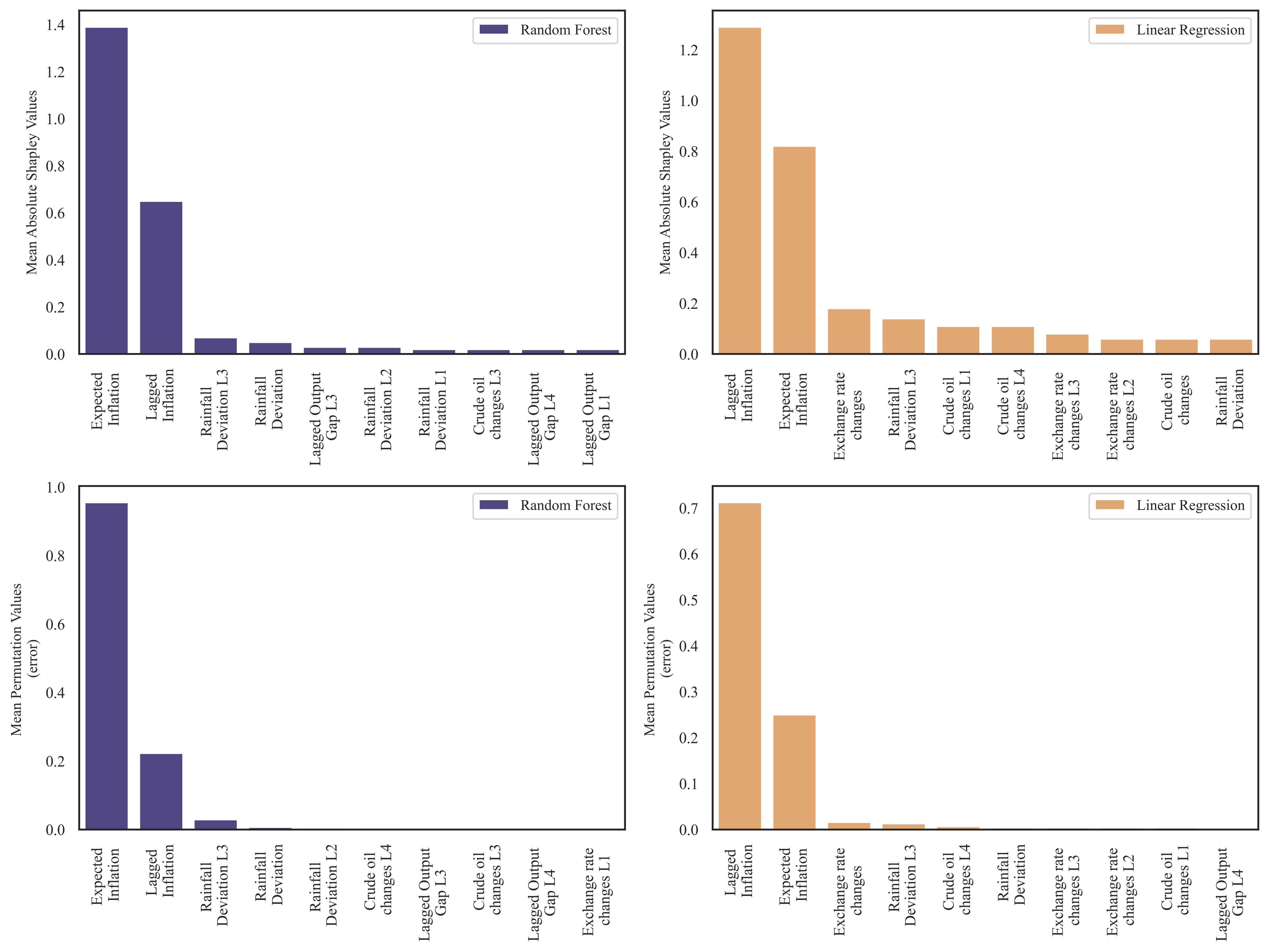}
   
    \label{fig:FI}
    
    \justifying{
    \footnotesize{{Note: Top panel plots the variable importance derived from Shapley Values, while the bottom panel plots the variable importance derived from Permutation Importance. Only the top 10 variables as per the specified criteria are selected for each model.}}}
\end{figure}

\textbf{Feature Importance} - Feature importance results using Shapley values and Permutation Importance are presented in Chart \ref{fig:FI}. Both models highlight inflation expectations and lagged inflation as significant drivers of inflation dynamics. However, the RF model places greater emphasis on expected inflation, as shown by its dominance in both Shapley values and permutation importance analyses. This suggests that the RF model is better equipped to capture forward-looking inflation expectations, which are critical in a policy-driven macroeconomic context. It also highlights the rising role of inflation expectations in determining the inflation process in India. 

In contrast, the LR model assigns more importance to lagged inflation, indicating that it relies more heavily on historical inflation data to make predictions. The output gap, a key element in understanding the inflation-economic slack relationship, is notably absent from the LR model's top 10 features but is captured by the RF model. This implies that the RF model is better suited to detect the potential nonlinear interactions between inflation and economic slack, a relationship that the LR model, due to its linear nature, fails to capture. Overall, the RF model demonstrates superior feature identification, particularly in its ability to weigh inflation expectations and the output gap more effectively than the LR model, aligning with its improved forecasting performance under the hybrid Phillips curve specification.

\begin{figure}[h!]
\caption{Partial Dependence Plots}
    \centering
    \includegraphics[width=0.95\linewidth]{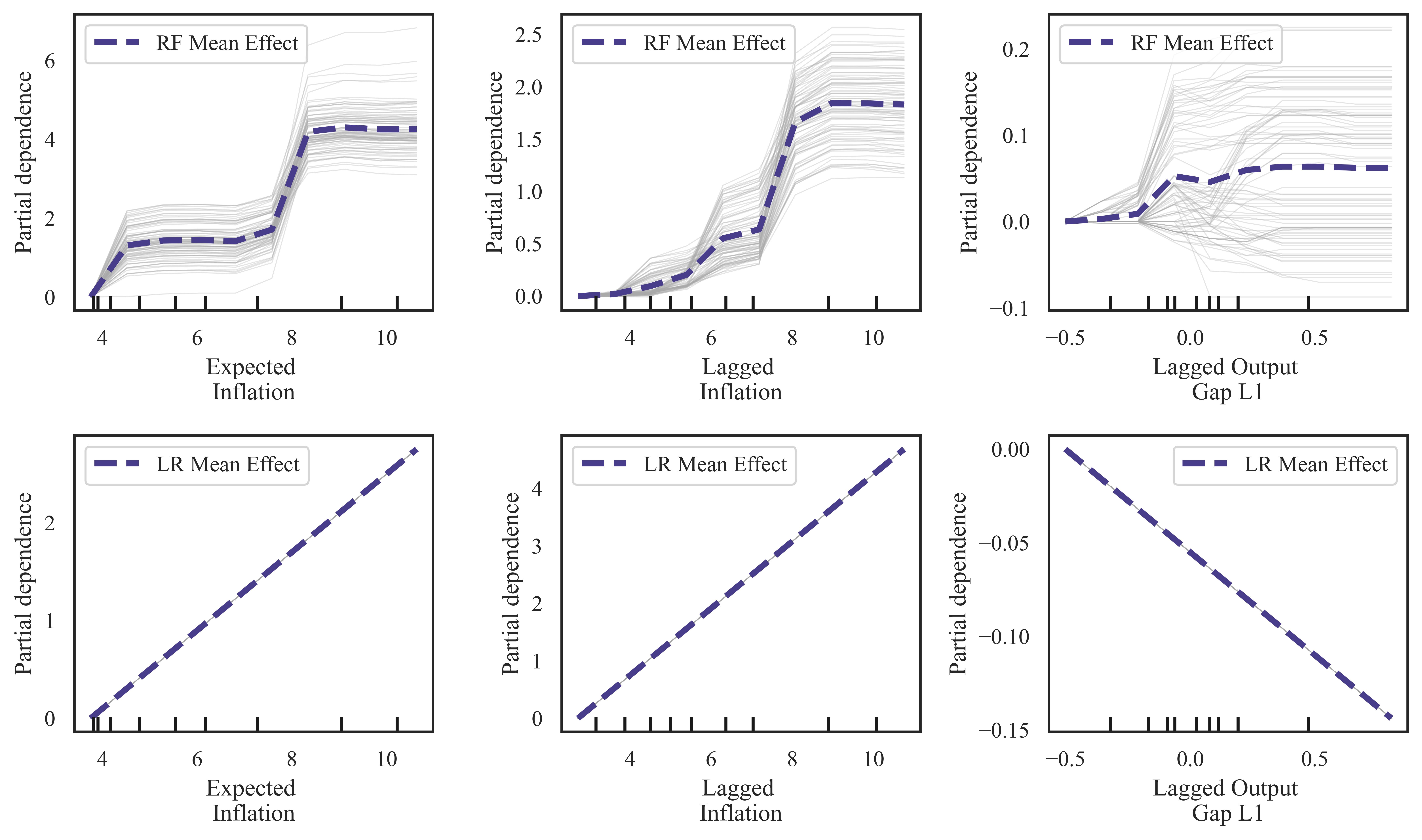}
    
    \label{fig:pdps}
    
    \justifying{
    \footnotesize{{Notes: i) The figure displays Partial Dependence Plots (PDPs) for three features: Expected Inflation, Lagged Inflation, and Lagged Output Gap. The dotted thick line plots the mean marginal effect while rest of the solid lines plot the Individual Conditional Expectation (ICE) for each individual row in the dataset. 
    
    ii) The top row corresponds to the Random Forest model, while the bottom row represents the Linear Regression model. PDPs are computed using grid resolution of 10. Grid resolution of 10 means that the feature is evaluated at 10 equally spaced points between the minimum and maximum values of the feature in the dataset. If you increase this number, the PDP would become more granular, capturing finer details of how the feature influences the predictions.}}}
\end{figure}

\textbf{Partial Dependence Plot} - The Partial Dependence Plots (PDPs) estimate the marginal effect of a given feature on the predicted outcome of a model, providing insight into the nature of the relationship between the predictor and the target variable. PDPs can reveal whether these relationships are linear, monotonic, or exhibit more complex patterns, allowing for a clearer understanding of modelled relationship. Chart \ref{fig:pdps} presents the PDPs for expected inflation, lagged inflation and output gap for both Random Forest and Linear Regression models. While the dotted blue line depicts the mean effect of the variable i.e., the partial dependence plot, the grey lines show the ICE plots for the given variable (feature). 

The RF model captures non-linear relationships in the data, as evidenced by the distinct, step-like patterns seen in its PDPs for expected inflation and lagged inflation. These results suggest that the RF model identifies \emph{thresholds} at which the impact of these variables on realized inflation shifts sharply, a characteristic that cannot be captured by the LR model. For instance, as inflation expectations increase, the prediction for inflation increases but this increase is subject to two thresholds – around 4.75 percent and 7.25 percent – suggesting that expected inflation has a nonlinear impact on actual inflation. Likewise, the ICE plots also presents a visual cue on the level of variation in model predictions when we fix the value of a feature. For example, lagged inflation of around 7.0 percent leads to a prediction that is 0.5 to 2.0 percent higher than the average prediction but such an effect is absent at lower values of lagged inflation. Similarly, the PDP for output gap, generated with the RF model, reflects a more complex interaction indicating potential non-linear relationship between output gap and inflation. Notably, the RF model correctly captures the positive relationship between the output gap and inflation, such that positive (negative) output gap leads to higher (lower) inflation, aligning with the Phillips curve theory and evidence. On the other hand, LR estimates a negative relationship in the sample.

The PDPs for the LR model, unsurprisingly, demonstrate linear relationship between inflation and all other variables in the model. The LR model assumes a constant, additive effect for each predictor, leading to straight-line PDPs that cannot account for any non-linear interactions in the data. This limitation underscores the fundamental difference between the two models: while LR provides a simple, interpretable structure, it fails to capture the more intricate relationships that are often present in macroeconomic data, such as those detected by the RF model. As such, the Random Forest's ability to identify non-linearities provides a more nuanced understanding of the inflation process.

\begin{figure}[h!]
 \caption{2-Dimensional Partial Dependence Plot}
    \centering
    \includegraphics[width=0.95\linewidth]{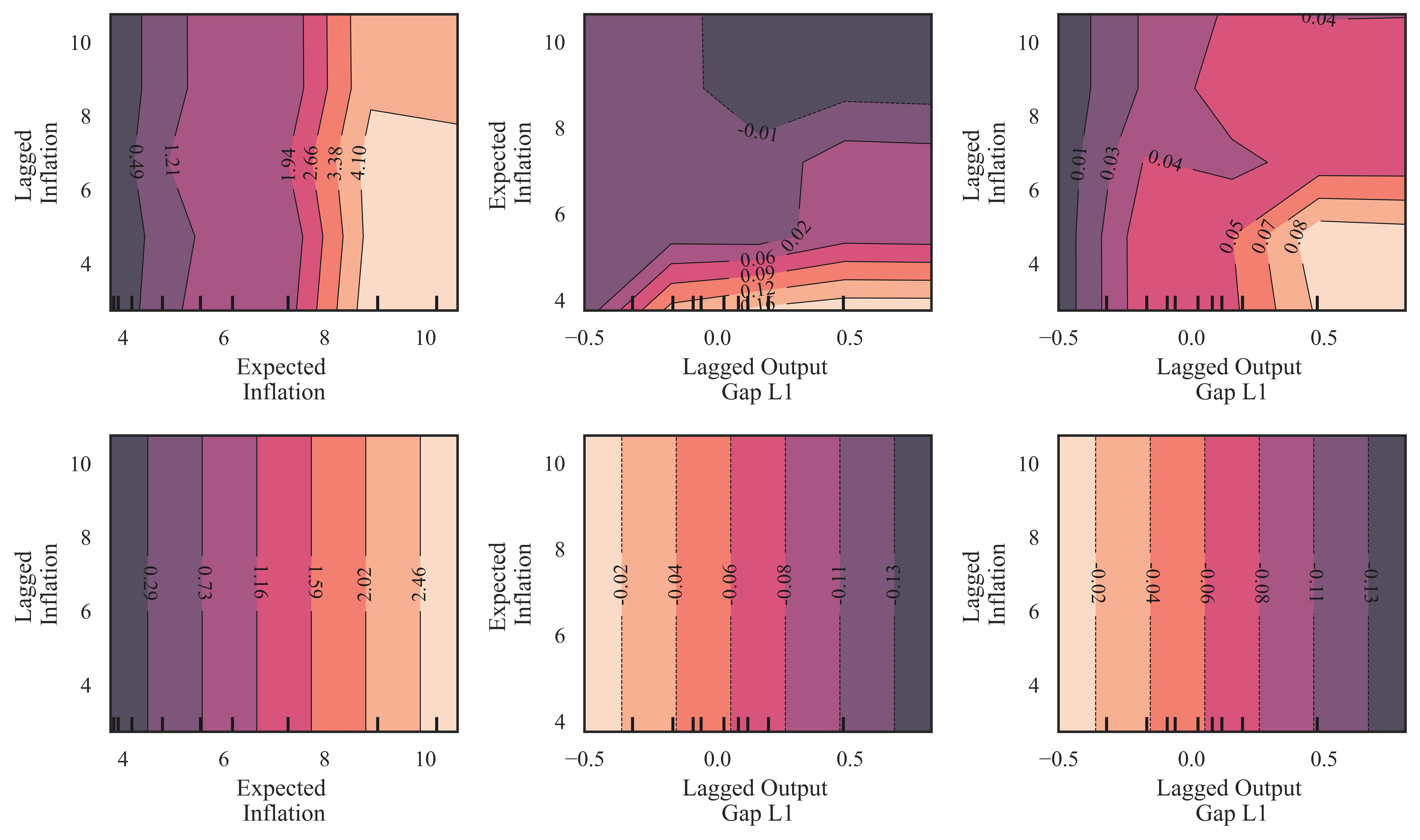}
    \label{fig:2dpdps}
    \justifying{
    \footnotesize{{Notes:The figure shows 2D Partial Dependence Plots (PDPs) for pairs of features: Expected Inflation, Lagged Inflation, and Lagged Output Gap;The top row represents the Random Forest model (non-linear interactions), and the bottom row shows the Linear Regression model (linear relationships); Grid resolution of 5 is used. Darker colours represent lower predictions, lighter colours higher. }}}
\end{figure}

The 2-dimensional Partial Dependence Plots (PDPs) provide insights into the interaction effects between key variables of interest, in this case, expected inflation, lagged inflation, and output gap on inflation forecasts. The Random Forest (top row) in Chart \ref{fig:2dpdps} captures complex, non-linear interactions between these variables. For instance, in the first plot (expected inflation vs. lagged inflation), inflation forecasts increase sharply when both expected and lagged inflation cross certain thresholds, indicating a compounding effect. This is possible when inflation expectations can get de-anchored in the face of already high inflation leading to an even higher surge in prices. Similarly, the second and third plots reveal non-linear interactions between output gap and both expected and lagged inflation, demonstrating that inflation is likely to be higher when the output gap is positive, particularly when combined with elevated inflation expectations.

In contrast, the Linear Regression (bottom row) shows strictly linear, additive interactions, as expected. The PDPs exhibit uniform, parallel contours, indicating that the LR model does not capture any interaction effects between the variables. Inflation predictions in the LR model are driven by independent, additive contributions of each variable, missing the non-linear dynamics that the Random Forest captures. This inability to model interactions may lead to less accurate predictions in complex macroeconomic environments, where the relationship between variables like inflation expectations and the output gap is inherently non-linear.

\begin{figure}[h!]
 \caption{Shapley Summary plot}
    \centering
    \includegraphics[scale = 0.65]{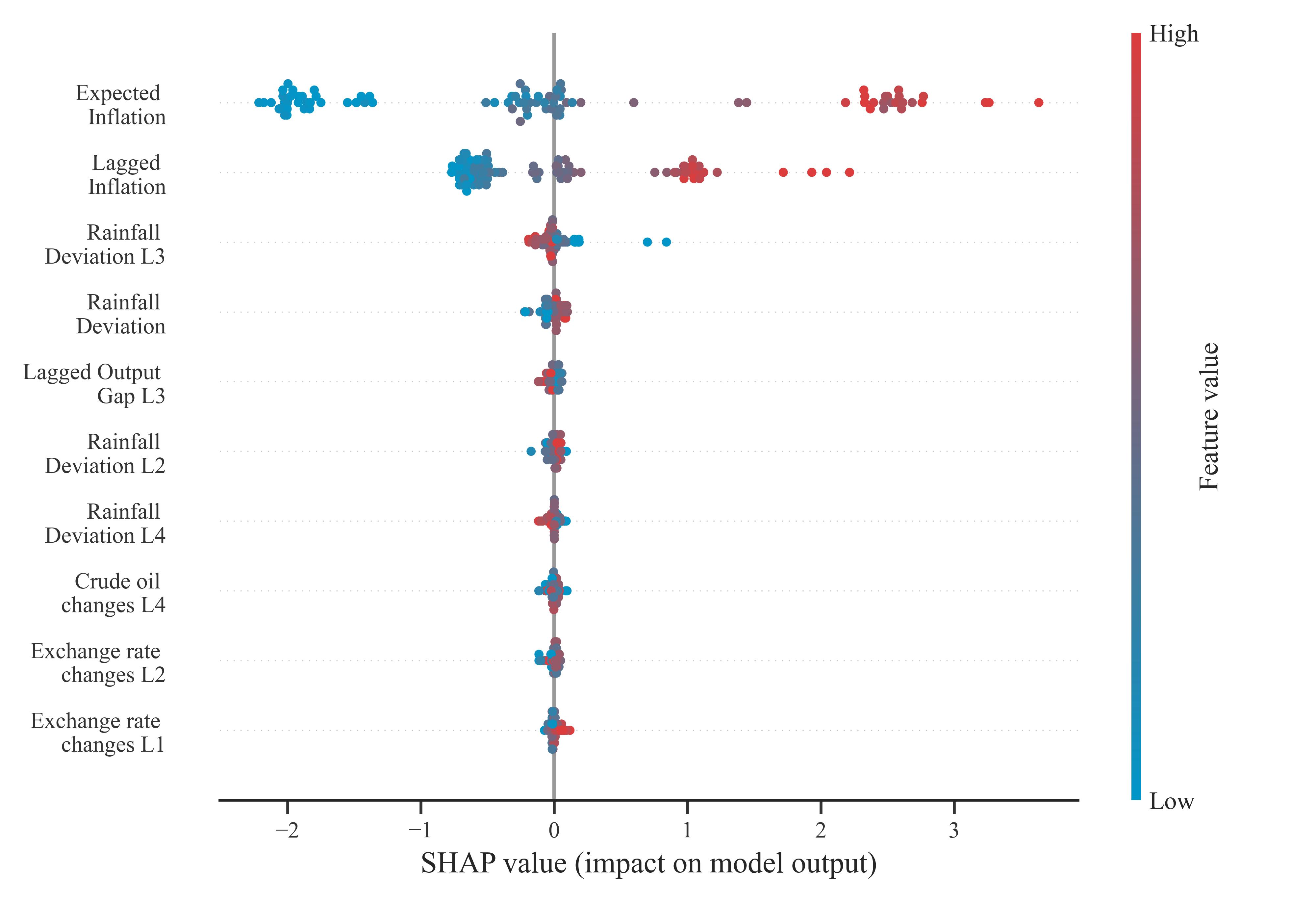}
    \label{fig:shapley_summary}

    \justifying{
    \footnotesize{{Note: The chart plots the variability of Shapley values for the distribution of feature values within our sample. Only the top 10 features are plotted.}}}
\end{figure}

\textbf{Shapley Values and Shapley Regression} - Chart \ref{fig:shapley_summary} shows the Shapley Summary Plot highlighting the contribution of the top-10 model features to the model's predictions. Each point on the plot represents a single observation's SHAP value for a given feature. The x-axis indicates the SHAP value, reflecting the feature's impact on the model's output, while the colour gradient represents feature values (blue for low, red for high).

Expected inflation and lagged inflation are the most influential features. Higher values of these features (shown in red) strongly increase inflation forecasts (positive SHAP values), while lower values (shown in blue) reduce inflation. The plot also highlights an asymmetric effect for both variables: high values of expected and lagged inflation tend to have a much larger positive impact on predictions compared to the negative impact from lower values. Other features, such as rainfall deviations and exchange rate changes, exhibit smaller and more symmetric influences on predictions, contributing to moderate positive or negative adjustments depending on their values.

This plot underscores the strong influence of inflation-related variables and the asymmetric nature of their impact on the model’s predictions, particularly at higher values of expected and lagged inflation.

\begin{figure}[h!]
    \centering
     \caption{Functional Form Learned by the Model}
    \includegraphics[width=0.95\linewidth]{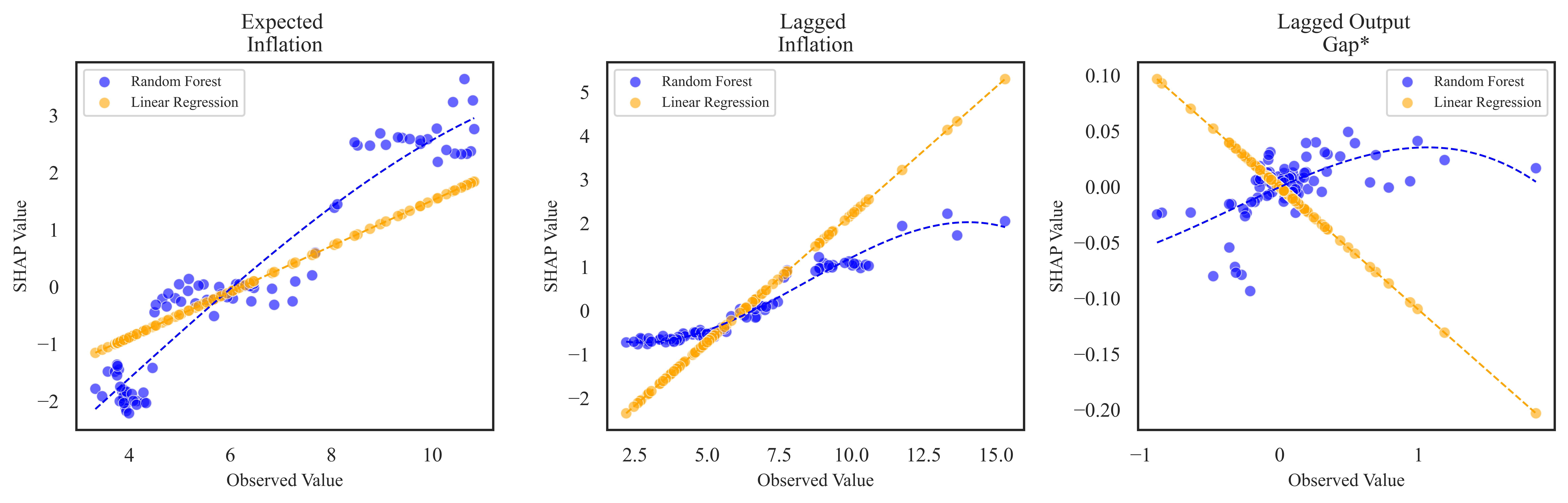}
   
    \label{fig:func_form}
    
    \justifying{
    \footnotesize{{Notes: i) This chart plots the Shapley values for the variable against its observed values for both Random Forest Model and a Linear Regression model. The dotted line through the scatter points is a 3rd-order best fit polynomial.
    
    ii) *: For output gap, a large outlier corresponding to June quarter of 2020 is omitted for clearer exposition.}}}
\end{figure}

Chart \ref{fig:func_form} illustrates how the Random Forest (RF) and Linear Regression (LR) model capture the relationships between inflation and its key determinants.

 Notably, being an ML model, the RF model is able to discern a non-linear pattern in the data. For expected inflation, both models exhibit a positive relationship, but the RF model is able to learn a non-linear relationship, wherein inflation forecasts become more sensitive to expected inflation with higher levels of future inflation. In other words, the higher values of inflation expectations have an asymmetrically higher impact on actual inflation \emph{vis-a-vis} lower levels of expected inflation. This contrasts with the constant, linear relationship estimated by the LR model. Similarly, in case of past inflation, the RF model again captures inherent non-linearities in the data, such that forecasts tend to plateau as past inflation stabilizes at lower values of inflation, says around the target rate of inflation. 

The most distinctive difference is seen in the relationship between inflation and output gap. The RF model identifies a complex, asymmetric relationship where positive and negative values of output gap influences inflation in different manner. This underscores that the level of inflation tends to be sensitive with respect to the level of slack in the economy. The LR model oversimplifies this, assuming a constant, linear relationship over time. 

Finally, Table \ref{tab:shapley_results} presents the Shapley Regression coefficients ($\beta_s^k$) and Shapley Share coefficient ($\Gamma$) for our ML-based Phillips curve model. The results provide insights into the alignment of key variables with the inflation as the target variable. The coefficients \( \beta_S \) measure the alignment of a variable with the target, where values close to 1 indicate perfect alignment. Values greater than 1 show underestimation of a variable’s effect, while values less than 1 show overestimation. The model’s hyperplane tilts towards a variable’s Shapley component when \(\beta_s^k\) is greater than 1 and away when it’s less than 1. The significance of a variable decreases as \( \beta_S \) approaches zero.

The shapely share coefficient  has three parts: the sign shows the direction of alignment, the magnitude measures how much of the model predictions is explained by the variable, and the significance level indicates confidence in rejecting the null hypothesis of zero/negative coefficients. This is similar to the interpretation of coefficients in a linear regression analysis. Expected inflation shows near-perfect alignment ($\beta_s^k$ = 1.033) and contributes significantly ($\Gamma$ = 0.159 or 15.9\%), meaning the model accurately captures its effect on inflation. Lagged inflation is also well-aligned ($\beta_s^k$ = 0.951 or 9.51\%), though slightly overestimated, with a meaningful contribution ($\Gamma$ = 0.075). In contrast, the model underestimates the impact of rainfall deviation ($\beta_s^k$ = 1.621) and the output gap ($\beta_s^k$ = 1.958), despite their statistical significance ($p < 0.01$). The positive contributions of the output gap ($\Gamma$ = 0.004) and rainfall ($\Gamma$ = 0.005) suggest their relevance for modelling inflation dynamics in India. However, the underestimation implies that the model may not fully capture their true influence in the given sample.

\begin{table}[h]
\centering
\caption{Shapley Regression Coefficients and Contributions}
\begin{tabular}{l|ll}
\hline
Variable Name & $\beta_k^s$ & $\Gamma$ \\
\hline
Intercept & 6.268*** & 0.719*** \\
Expected Inflation & 1.033*** & 0.159*** \\
Lagged Inflation & 0.951*** & 0.075*** \\
Rainfall Deviation L3 & 1.621*** & -0.008*** \\
Rainfall Deviation & 1.422** & 0.005** \\
Lagged Output Gap L3 & 1.958*** & 0.004*** \\
Rainfall Deviation L2 & 1.01 & -0.003 \\
Rainfall Deviation L4 & 1.38 & -0.003 \\
Crude oil change L4& 1.934 & -0.003 \\
Exchange rate change L2& 1.23 & -0.002 \\
Exchange rate change L1& 2.368* & 0.002* \\
Crude oil change& 3.560*** & 0.002*** \\
Crude oil change L1& 2.940*** & 0.002*** \\
Rainfall Deviation L1 & 2.949** & 0.002** \\
Crude oil change L2& 2.411* & 0.002* \\
Lagged Output Gap L1 & 0.938 & -0.002 \\
Lagged Output Gap L4 & 3.644 & -0.002 \\
Crude oil change L3& 0.804 & -0.002 \\
Exchange rate change& 5.457*** & -0.001*** \\
Lagged Output Gap L2 & 4.615** & 0.001** \\
Exchange rate change L3& 1.845 & -0.001 \\
Exchange rate change L4& 8.250** & -0.001** \\
\hline
        \multicolumn{3}{p{0.5\textwidth}}{\small{Note: i) The Shapley regression is estimated using the Shapley values for the entire sample.  
        
        ii) Significance levels: *$p<0.1$, **$p<0.05$, ***$p<0.01$}}
\end{tabular}
\label{tab:shapley_results}

\end{table}

\newpage
\section{Robustness Checks and Additional Results}

\subsection{Robustness Checks}\label{sec:robustness}

To ensure the robustness of our findings, we expand our analysis in several ways. First, to evaluate the predictive accuracy of ML models, we test them against a broader set of baseline models. These baselines include univariate autoregressive (AR) models and multivariate vector autoregression (VAR) models. The results, presented in Table \ref{tab: univ_models}, indicate that RF consistently outperforms other univariate models for two and three quarter-ahead forecasts, achieving the lowest RMSE values of 1.286 and 1.544, respectively. In case of four quarter-ahead forecasts, RF remains highly competitive, with an RMSE of 1.813, closely aligning with the performance of AR(4) (1.803). VAR-based forecasts perform poorly across horizons. It should be noted that AR and VAR forecasts were generated using a recursive forecasting approach, whereas the RF model employed a direct forecasting methodology. This methodological distinction further underscores the superior performance of RF, demonstrating its effectiveness in capturing complex inflation dynamics across varying forecast horizons.
\begin{table}[h!]
\centering
\caption{RMSE for various models across forecast horizons}
\begin{tabular}{l|cccc}
\hline
\textbf{Model} & \textbf{Q1} & \textbf{Q2} & \textbf{Q3} & \textbf{Q4} \\
\hline
AR(1)    & 0.958 & 1.401 & 1.651 & 1.807 \\
AR(4)    & \textbf{0.940} & 1.392 & 1.637 & \textbf{1.803} \\
Random Forest & 0.949 & \textbf{1.286} & \textbf{1.544} & 1.813 \\
\hline
Multivariate models\\
\hline
VAR(1)  & 0.966 & 1.415 & 1.698 & 1.988 \\
VAR(4)  & 1.299 & 1.896 & 2.205 & 2.370 \\
\hline
\multicolumn{5}{p{.6\textwidth}}{\footnotesize{Note: i) The table presents the results of a horse-race between several models. Numbers in parenthesis highlight the lag order of the model.

ii) The VAR model is a reduced-form version of a simple open-economy model with output gap, inflation, exchange rate and 3-month Treasury Bill rate.}}
\end{tabular}
\label{tab: univ_models}
\end{table}

\begin{figure}[h!]
    \centering
    \caption{Forecasts Comparison}
    \includegraphics[scale = 0.7]{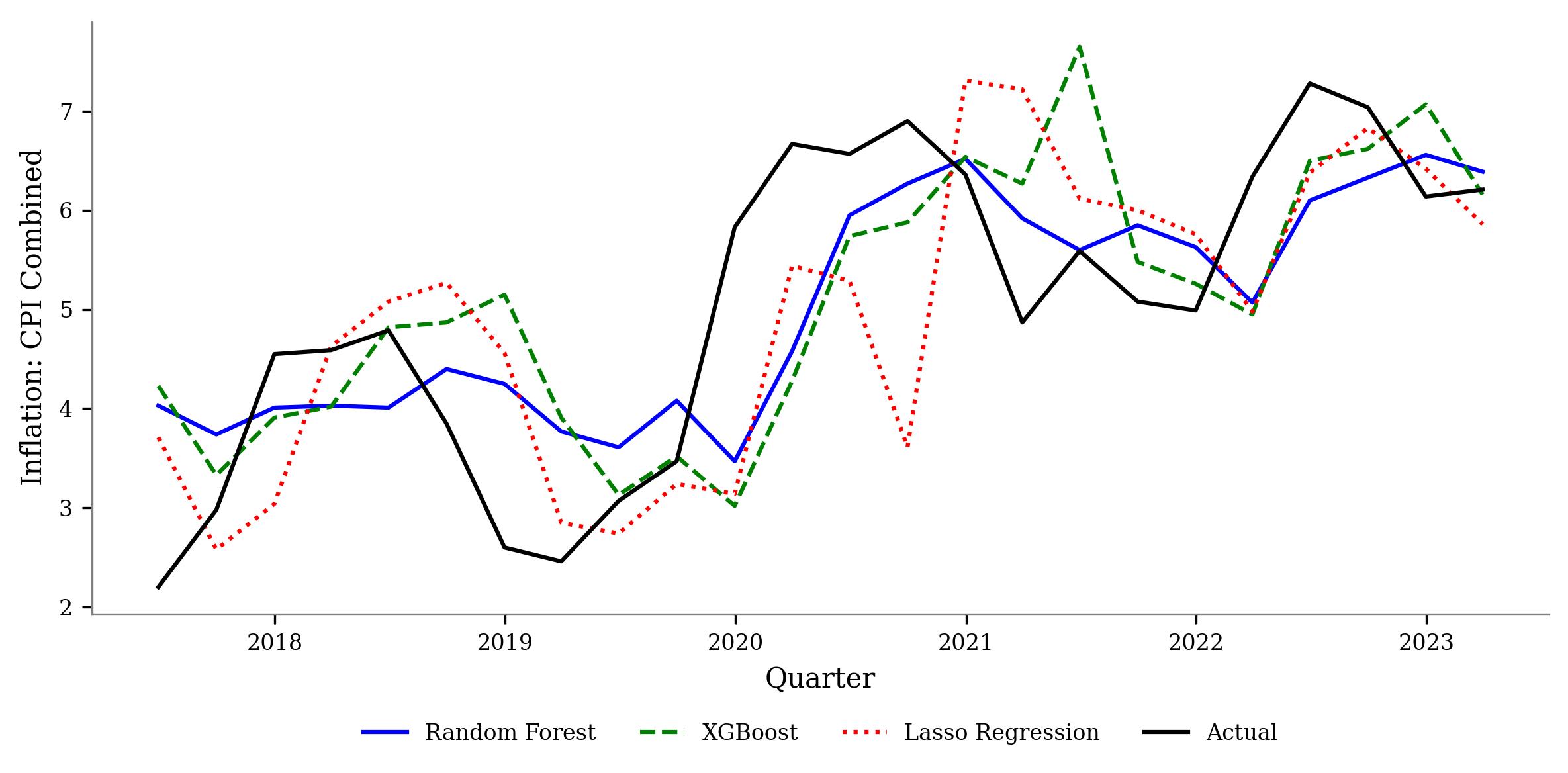}
    \label{fig:forecast_errbands}
     
\justifying{
\footnotesize{{Note: The chart plots the one-quarter ahead forecasts from the best performing linear and ML model against actual inflation and a Random walk baseline for the hybrid Phillips curve specification.}}}
\end{figure}

Chart \ref{fig:forecast_errbands} compares the forecasting performance of Random Forest, XGBoost, and Lasso Regression for predicting one-quarter ahead inflation.

\begin{figure}[ht!]
    \centering
    \caption{MCB Plot for RMSE Metric}
    \includegraphics[width=0.99\linewidth]{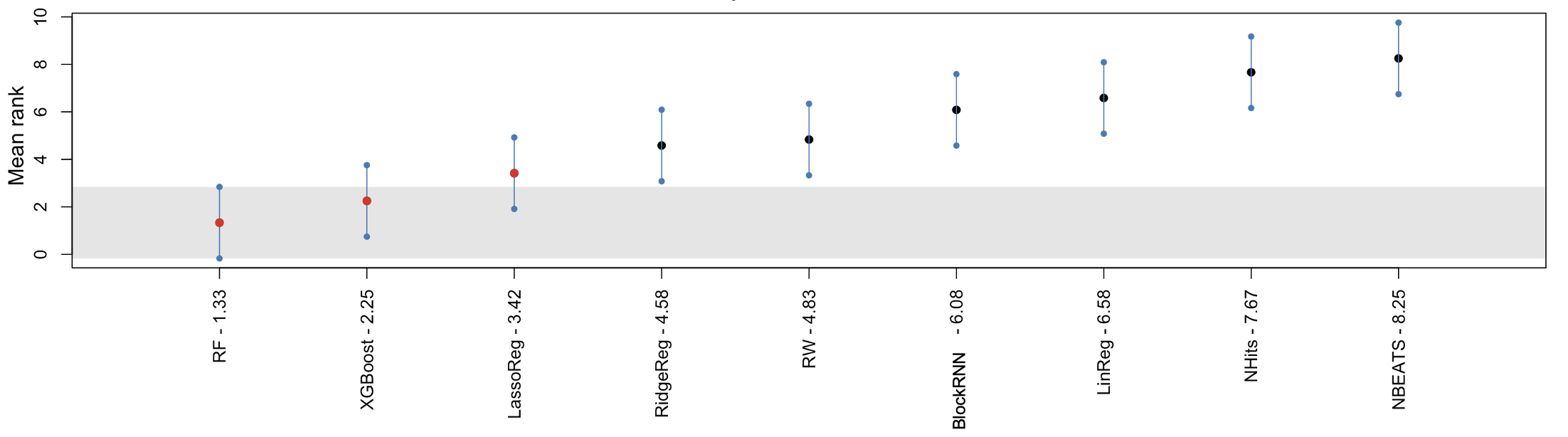}
   \justifying{
   \footnotesize{{Notes: The plot presents visualization of the multiple comparisons with the best (MCB) analysis across all four forecast horizons (Q1, Q2, Q3 and Q4) and for all the 3 Phillips curve regimes. In the figure, for example, ‘RF - 1.33’ means
that the average rank of the algorithm RF based on the RMSE error metric is 1.33; the same explanation applies to other algorithms.
    }}}
    \label{fig: MCB Plot}
\end{figure}

Second, to further evaluate the relative effectiveness of ML models, we employ a model-agnostic Multiple Comparisons with the Best (MCB) procedure (\citealp{koning2005m3}). The MCB test ranks models based on their ex-ante accuracy across datasets and determines statistical significance using critical distances (CD). This non-parametric approach identifies the model with the lowest mean rank as the "best" performer and calculates CDs as $\Theta_{\alpha} \sqrt{\frac{\mathcal{M}\left(\mathcal{M} + 1 \right)}{6 \mathcal{D}}}$, where $\Theta_{\alpha}$ is the critical value from the Tukey distribution at level $\alpha$. Chart \ref{fig: MCB Plot} summarizes the results of the MCB test for the RMSE metric. The results demonstrate that the Random Forest (RF) model is able to achieve the lowest mean rank of 1.33, establishing itself as the top-performing forecasting method. This is followed by XGBoost with a mean rank of 2.25. Other models, including Lasso Regression (3.42), Ridge Regression (4.58), and Random Walk (4.83), trail further behind. Interestingly, deep learning models such as BlockRNN, NHits, and NBEATS rank even lower with mean ranks of 6.08, 7.67, and 8.25, respectively. The shaded region in the figure represents the critical distance for the RF model serving as the benchmark for statistical significance. Models with intervals that overlap this critical distance, such as the XGBoost model, exhibit performance close to RF. On the other hand, other models including neural networks fall significantly short. This further underlines the superiority of RF in forecasting CPI inflation over both short-term and long-term horizons. 

Third, in general, machine learning models demonstrate superior adaptability and performance in rapidly evolving environments. To 
examine this, we train and measure the forecasting performance of our ML models across different sample periods. This is pertinent in the current context given the macroeconomic volatility induced by the Covid-19 pandemic. When we examine forecasts in the pre- and post-Covid period, a clear pattern emerges (see Table \ref{tab: rmse_models}). During the relatively stable pre-Covid period, characterized by moderate inflation and steady economic growth, all models perform well. In this case, linear approaches like the LASSO model prove adequate. However, the sample period after the outbreak of Covid-19 pandemic was marked by large-scale disruptions, supply chain bottlenecks, evolving consumer behaviour, and geopolitical shocks such as the Russia-Ukraine conflict. In this environment, ML models like the Random Forest and XGBoost outperform LASSO Regression by effectively capturing nonlinear patterns and rapid shifts in inflation dynamics. These findings underscore that flexible machine learning models can provide better forecasts to navigate periods of high volatility and complex structural changes. 

\begin{table}[h!]
\centering
\caption{RMSE comparison across time periods}
\begin{tabular}{l|c|c}
\hline
\textbf{Model} & \textbf{RMSE (Pre-Covid19)} & \textbf{RMSE (Post-Covid19)}\\
\hline
Lasso Regression & 1.287 & 1.387 \\
Random Forest    & \textbf{1.210} & \textbf{0.920} \\
XGBoost          & 1.431 & 1.160 \\
\hline
\multicolumn{3}{p{.8\textwidth}}{\footnotesize{Note: The table presents the Root Mean Squared Error (RMSE) for Lasso Regression, Random Forest, and XGBoost models for forecasting inflation. The periods are divided into pre-COVID (up to December 2019) and post-COVID (from March 2020). Lower RMSE values indicate better model performance.}}
\end{tabular}
\label{tab: rmse_models}
\end{table}

Fourth, in order to analyze the forecasting performance of our best performing model across time in a more formal manner, we employ the Giacomini and Rossi (GR) test to assess the accuracy of the RF model relative to the RW and XGBoost model. Proposed by \citet{giacomini2010forecast}, the GR test provides a formal statistical approach to assess forecast accuracy over time by employing a rolling evaluation windows. Thus, the test allows a researcher to identify if one model consistently outperforms other models across different time periods or if performance fluctuates between models. For brevity, the analysis is restricted to a four quarter-ahead forecast horizon. The null hypothesis of the test assumes equal performance between models across all time points against the alternative which identifies specific instances of divergence. Unlike traditional forecast evaluation tests, such as those proposed by \citet{diebold2002comparing} and \citet{rossi2016forecast}, the GR test is uniquely suited for analyzing time-varying model performance under unstable conditions.

\begin{figure}[ht!]
    \centering
    \caption{Giacomini and Rossi test: Results}
    \includegraphics[width=0.95\linewidth]{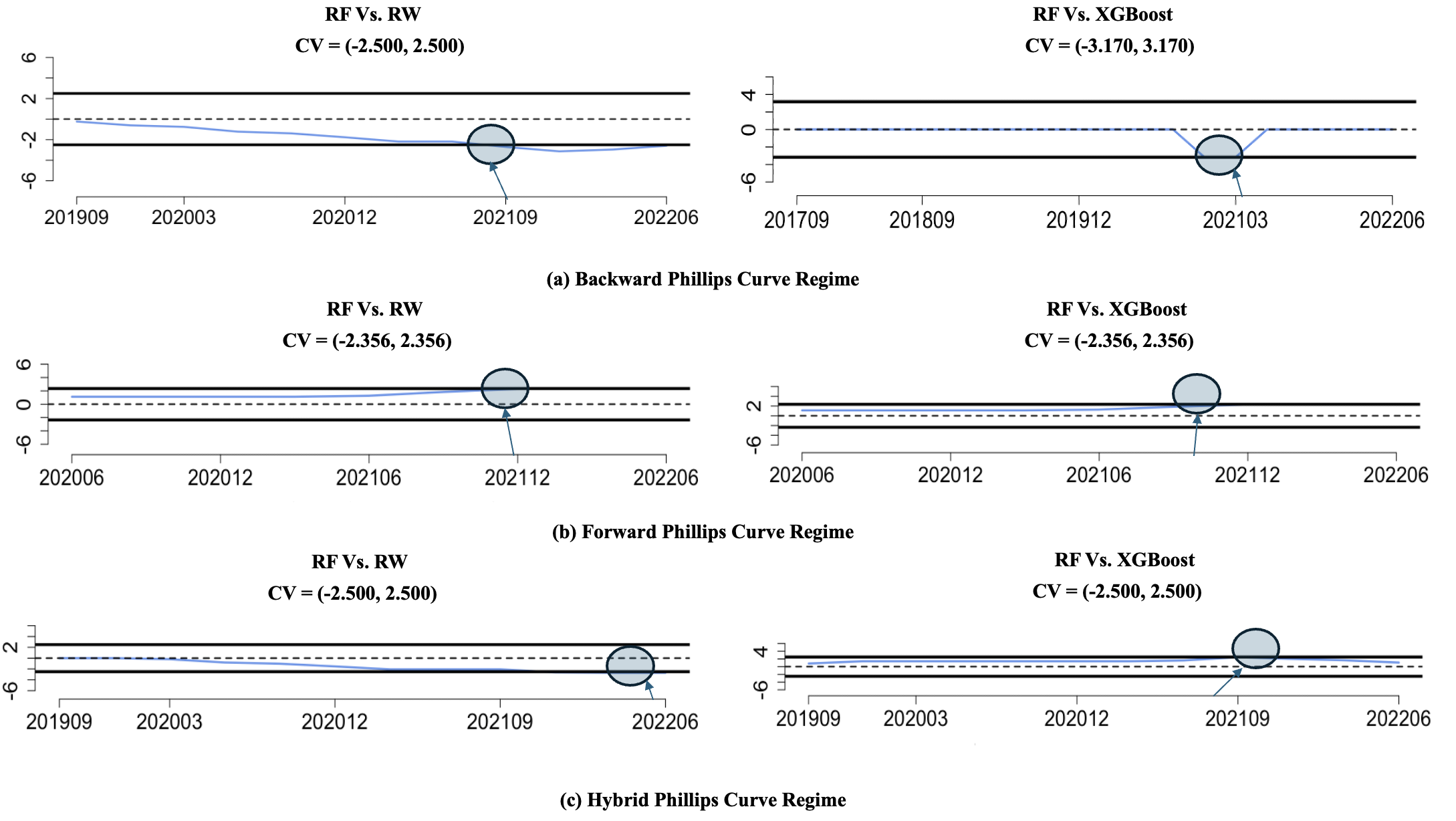}
   \justifying
   \footnotesize{Notes: The plots present visualization Results of the GR test (Giacomini \& Rossi test) of the forecasting ability of RF with baselines (RW (left) and XGBoost (right)) for 4-quarter ahead forecasts with ($\alpha$ = 0.90) for 3 Phillips Curve regimes. Points outside the CV are marked under the blue-shaded circles, representing that RF algorithm and the other two top-performing benchmarks differ significantly.}
    \label{fig: GR tests}
\end{figure}

The results of the GR test are shown in Chart \ref{fig: GR tests}. The chart shows the critical values (CVs) represented by thick black lines to delineate thresholds for statistical significance. Parameter $\mu$ defines the rolling window size relative to the evaluation sample. As the results showcase, the RF model exhibits consistent improvements over baseline models, with notable divergence observed in the post-Covid sample marked by heightened economic uncertainty and structural disruptions. For the backward-looking PC specification, RF model significantly surpasses both XGBoost and RW in 2021. A similar pattern emerges in case of both the forward-looking and hybrid PC models. The increased variability in the relative performance of the RF model during the post-pandemic period highlights the impact of structural changes on the underlying data relationships. These findings further confirms the robustness and superior predictive accuracy of ML models, particularly in the volatile post-pandemic environment. 

Finally, we explore three methodological variations within the Phillips curve framework: (a) employing the HP filter to estimate trend inflation and the output gap, (b) constructing a composite measure of supply shocks using the principal components approach (PCA) to account for aggregate supply-side dynamics, and (c) integrating alternative measures of inflation expectations derived from internet search volume data, as proposed by \citet{bicchal2019rationality}. These variations allow us to rigorously assess how different proxies and assumptions related to inflation drivers influence forecasting outcomes. Across all specifications, ML models consistently demonstrate strong predictive performance, with Random Forest and XGBoost maintaining competitiveness across scenarios. LASSO Regression also performs well in specific cases, showcasing its potential in certain contexts. Detailed results of these robustness checks are provided in Appendix A (Tables A5 through A7), further substantiating the efficacy of ML models under varying assumptions.

\subsection{Uncertainty Quantification through Conformal Prediction Intervals (CPI)}
A critical aspect of forecasting, particularly in complex and non-linear settings, is quantifying the uncertainty associated with predictions. In this section, we leverage the conformal prediction interval (CPI) framework to assess the uncertainty of forecasts generated across the three Phillips Curve specifications—backward, forward, and hybrid. The conformal prediction approach, first introduced by \citet{vovk2005algorithmic}, provides a non-parametric methodology for constructing robust prediction intervals. Unlike traditional confidence intervals, which primarily quantify the uncertainty of model parameters in a parametric setup, conformal prediction intervals are designed to capture the uncertainty of the forecast or predictive outcome, making them highly relevant for forecasting applications. For our study, we construct CPIs using the sequential nature of time series data that leverages its temporal ordering to build prediction intervals (\citealp{sengupta2024forecasting}).

Consider a training data set $\left\{Y_t, \bar{X}_t \right\}_{t=1}^N$ where $Y_t$ represents the target variable to be forecasted, and $\bar{X}_t$ denotes the set of predictor variables or features. To quantify the uncertainty associated with predictions, we fit the Random Forest (RF) model alongside an uncertainty model $\widehat{\Xi}$ on $\bar{X}_t$ which maps $\bar{X}_t$ to a scalar uncertainty measure. The conformal score $\mathcal{S}_{t}$ can be calculated as:

\begin{equation}
        \mathcal{S}_{t} = \frac{\left|Y_{t} - \operatorname{RF}\left(\bar{X}_{t} \right)\right|}{\widehat{\Xi}(\bar{X}_{t})}
\end{equation}

Given the sequential structure inherent in time-series data, conformal scores are adjusted using weights to reflect temporal dependencies. Specifically, a fixed $\kappa$-sized sliding window is applied, where the weights $\omega_{t^{'}}$ are defined as $\omega_{t^{'}} = \mathbb{1}\{ t^{'} \geq t - \ \kappa), \forall\ t^{'} < t$. The time-adjusted quantile values $\widehat{\mathbb{Q}}_t$ are then computed as:
\begin{equation}
    \widehat{\mathbb{Q}}_{t} = inf\left\{ q:\ \frac{1}{\min\left( \kappa,t^{'} - 1 \right) + 1}\sum_{t^{'} = 1}^{t - 1}{\mathcal{S}_{t^{'}}\mathbb{1}\left(t^{'} \geq t - \ \kappa \right) \geq 1 - \alpha}\right\}
\end{equation}

These quantiles are critical in constructing prediction intervals that adjust dynamically for data heterogeneity and reflect the uncertainty around predictions. Using the weight-adjusted quantiles $\widehat{\mathbb{Q}}_t$, the conformalised prediction interval for each time step $t$ is given by:
\begin{equation}
    \mathbb{C}\left(\bar{X}_{t} \right) = \left[\operatorname{RF}\left(\bar{X}_{t} \right) - {\widehat{\mathbb{Q}}}_{t} \widehat{\Xi}\left(\bar{X}_{t} \right), \operatorname{RF}\left( \bar{X}_{t} \right) + {\widehat{\mathbb{Q}}}_{t}\ \widehat{\Xi}\left(\bar{X}_{t} \right)\right].
\end{equation}

% ...
\begin{figure}[ht!]
    \centering
    \caption{4Q Ahead Forecast Performance and Conformal Prediction Intervals}
    \includegraphics[width=0.999\linewidth]{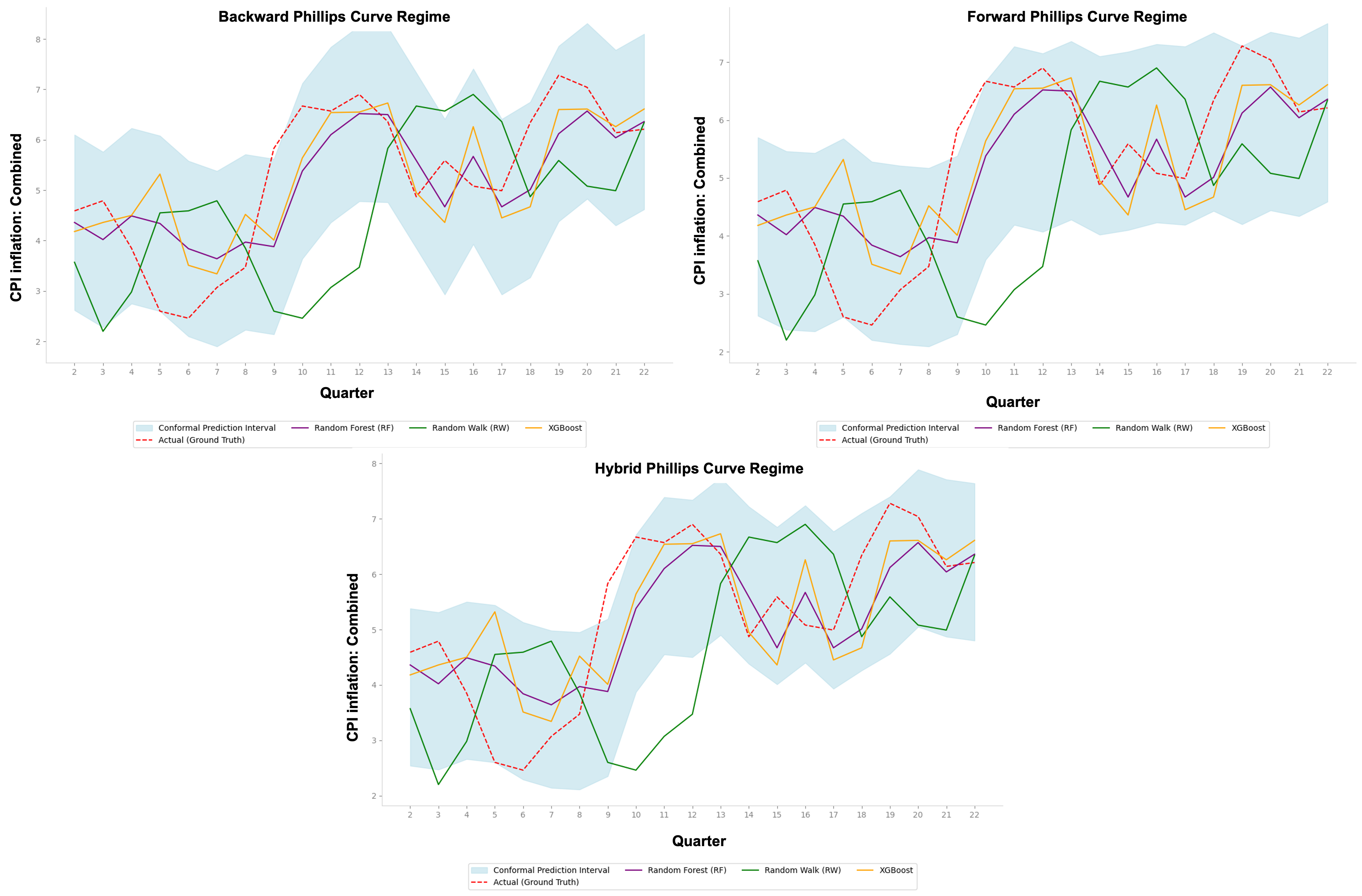}
   \justifying{
   \footnotesize{{Notes: The plot shows the ground truth CPI inflation:combined data (red dots), 4Q-ahead point forecasts generated by Random Forest(RF) (purple line), Random Walk(RW) (green line), XGBoost (yellow line), and the conformal prediction interval of RF (light blue shaded) for 3 Phillips curve regime. 
    }}}
    \label{fig: CPI}
\end{figure}
We analyze CPIs derived from the Random Forest (RF) model for 4-quarter-ahead forecasts of headline inflation over 21 quarters from 2017:Q2 to 2022:Q2. The RF model is selected as the best-performing approach based on the MCB test discussed earlier. It demonstrated superior predictive accuracy compared to both linear models (linear regression, lasso regression, and ridge regression) and other non-linear models (XGBoost, NBeats, NHits, and BlockRNN). Chart \ref{fig: CPI} illustrates the CPIs produced by the RF model along with point forecasts generated using the RF model, the XGBoost model (second-best) , and the Random Walk (RW) model. Including XGBoost forecasts provides a benchmark for evaluating the RF model against a competitive alternative, while the RW model serves as a baseline to assess performance against a simplistic forecasting strategy. 

Though it often serves as a simplistic yet effective baseline, the RW model fails to capture nuanced variations in headline inflation, particularly during periods of high volatility. This limitation is evident in its inability to align closely with the ground truth data across all three model specifications. In contrast, the RF model not only demonstrates competitive point-forecasting capabilities but also offers a rigorous mechanism for quantifying predictive uncertainty through CPIs. A notable feature of CPIs is the variability in width across the three specifications reflecting differences in forecast uncertainty. For instance, the average interval width is approximately 3.48 for the backward-looking PC, 3.08 for the forward-looking PC (FPC) regime, and 2.84 for the hybrid PC. This suggests that forecasts from the RF model exhibit lower uncertainty in a hybrid specification. This is likely because the hybrid model combines both backward- and forward-looking elements that may be better at capturing the structural dynamics of inflation.

Therefore, conformal prediction approach allows us to quantify forecast uncertainty in our models and assess their robustness over a multi-step ahead forecast horizon. Being a non-parametric, model-agnostic method for quantifying forecast uncertainty, CPIs can be a valuable addition to the forecasting toolkit, especially in the context of non-linear and complex economic relationships such as that encapsulated in the Phillips Curve framework for India.

\clearpage
\section{Conclusion} \label{sec:conclusion}
Machine learning (ML) and artificial intelligence (AI) have been around for decades, but they have only recently achieved ubiquity due to rapid advances in computing power and data availability. ML/AI algorithms excel at prediction tasks, but they have traditionally struggled to explain their predictions. This has been a major reason for their slow adoption in economic policy.

Therefore, in this paper, we propose machine learning (ML) models as a credible alternative to classical econometric approaches commonly used in empirical macroeconomics, specifically within the context of inflation modelling and forecasting. We employ a New Keynesian Phillips Curve (NKPC) framework to examine inflation dynamics and its determinants in the context of India. Our findings suggest that ML techniques not only outperform conventional linear models but also offer a nuanced approach for analysing complex, nonlinear relationships in the data through explainable machine learning methods.

While ML has traditionally been criticised for its \emph{black-box} nature, the application of techniques such as feature importance, partial dependence plots, and Shapley values allow us to peep into this black box, providing interpretable insights into the structural determinants of inflation in India. These tools enable a deeper understanding of the inflationary process, revealing the importance of forward-looking elements i.e., inflation expectations and the presence of threshold effects in variables such as the output gap. This allows us to model the non-linearities and structural breaks that conventional models often fail to capture.

However, certain limitations in our analysis should be acknowledged. First, ML models are inherently data-intensive, and our study could benefit from a larger sample size to enhance robustness and potentially uncover further non-linearities in the data. Second, while our results show that expected inflation is a crucial determinant of realised inflation, the choice of the inflation expectations measure remains open to debate. Using alternative measures, such as household or professional forecasts, may yield different outcomes. Similarly, the output gap, which proxies marginal cost in the Phillips curve, is a latent and often imprecisely estimated variable. Variations in its measurement could influence the model's conclusions. Lastly, for large datasets both training ML and Deep learning models as well as calculating shapley values discussed in the paper may be prohibitively compute intensive which the researcher/practitioner must keep in mind while trying to implement this framework in other prediction tasks.

Notwithstanding its limitations, this study demonstrates the potential of machine learning techniques for modeling complex economic relationships, particularly in settings characterized by structural breaks and volatility, such as inflation forecasting in developing economies. The application of Shapley values facilitates the disaggregation of next-period inflation into supply-side (e.g., weather or oil shocks) and demand-side (e.g., output gap) components which can have important implications for monetary policy. This research establishes a foundation for subsequent investigations into macroeconomic relationships through machine learning, addressing the constraints of traditional models in economic forecasting and policy formulation. Future studies may assess the generalizability of these findings across emerging market economies and extend similar methodologies to other macroeconomic variables.  

\clearpage
\onehalfspacing
\setlength\bibsep{3pt}
\bibliographystyle{jss2}
\bibliography{biblio}

\begin{thebibliography}{92}
\newcommand{\bibenquote}[1]{``#1''}
\providecommand{\natexlab}[1]{#1}
\providecommand{\url}[1]{\texttt{#1}}
\providecommand{\urlprefix}{URL }
\expandafter\ifx\csname urlstyle\endcsname\relax
  \providecommand{\doi}[1]{doi:\discretionary{}{}{}#1}\else
  \providecommand{\doi}{doi:\discretionary{}{}{}\begingroup \urlstyle{rm}\Url}\fi
\providecommand{\eprint}[2][]{\url{#2}}

\bibitem[{Almosova and Andresen(2023)}]{almosova2023nonlinear}
Almosova A, Andresen N (2023).
\newblock \bibenquote{Nonlinear inflation forecasting with recurrent neural networks.}
\newblock \emph{Journal of Forecasting}, \textbf{42}(2), 240--259.

\bibitem[{Athey and Imbens(2019)}]{athey2019machine}
Athey S, Imbens GW (2019).
\newblock \bibenquote{Machine learning methods that economists should know about.}
\newblock \emph{Annual Review of Economics}, \textbf{11}(1), 685--725.

\bibitem[{Ball \emph{et~al.}(2016)Ball, Chari, and Mishra}]{ball2016understanding}
Ball L, Chari A, Mishra P (2016).
\newblock \bibenquote{Understanding inflation in India.}
\newblock \emph{Technical Report w22948}, National Bureau of Economic Research.

\bibitem[{Ball and Mazumder(2019)}]{ball2019phillips}
Ball L, Mazumder S (2019).
\newblock \bibenquote{A Phillips curve with anchored expectations and short‐term unemployment.}
\newblock \emph{Journal of Money, Credit and Banking}, \textbf{51}(1), 111--137.

\bibitem[{Ball \emph{et~al.}(2011)Ball, Mazumder, Dynan, and Stock}]{ball2011inflation}
Ball L, Mazumder S, Dynan K, Stock JH (2011).
\newblock \bibenquote{Inflation Dynamics and the Great Recession/Comments and Discussion.}
\newblock \emph{Brookings Papers on Economic Activity}, pp. 337--.

\bibitem[{Behera \emph{et~al.}(2018)Behera, Wahi, and Kapur}]{behera2018phillips}
Behera H, Wahi G, Kapur M (2018).
\newblock \bibenquote{Phillips curve relationship in an emerging economy: Evidence from India.}
\newblock \emph{Economic Analysis and Policy}, \textbf{59}, 116--126.

\bibitem[{Bernanke(2003)}]{bernanke_2003}
Bernanke B (2003).
\newblock \bibenquote{Constrained Discretion and Monetary Policy.}
\newblock Remarks Before the Money Marketeers of New York University.

\bibitem[{Bicchal and Durai(2019)}]{bicchal2019rationality}
Bicchal M, Durai SR (2019).
\newblock \bibenquote{Rationality of inflation expectations: an interpretation of Google Trends data.}
\newblock \emph{Macroeconomics and Finance in Emerging Market Economies}, \textbf{12}(3), 229--239.

\bibitem[{Bobeica and Hartwig(2023)}]{bobeica2023covid}
Bobeica E, Hartwig B (2023).
\newblock \bibenquote{The COVID-19 shock and challenges for inflation modelling.}
\newblock \emph{International journal of forecasting}, \textbf{39}(1), 519--539.

\bibitem[{Breiman(2001)}]{breiman2001random}
Breiman L (2001).
\newblock \bibenquote{Random forests.}
\newblock \emph{Machine learning}, \textbf{45}(1), 5--32.
\newblock Retrieved from: \url{https://link.springer.com/content/pdf/10.1023/A:1010933404324.pdf}.

\bibitem[{Buckmann and Joseph(2023)}]{buckmann2023interpretable}
Buckmann M, Joseph A (2023).
\newblock \bibenquote{An Interpretable Machine Learning Workflow with an Application to Economic Forecasting.}
\newblock \emph{International Journal of Central Banking}.

\bibitem[{Calvo(1983{\natexlab{a}})}]{CALVO1983383}
Calvo GA (1983{\natexlab{a}}).
\newblock \bibenquote{Staggered prices in a utility-maximizing framework.}
\newblock \emph{Journal of Monetary Economics}, \textbf{12}(3), 383--398.
\newblock ISSN 0304-3932.
\newblock \doi{https://doi.org/10.1016/0304-3932(83)90060-0}.
\newblock \urlprefix\url{https://www.sciencedirect.com/science/article/pii/0304393283900600}.

\bibitem[{Calvo(1983{\natexlab{b}})}]{calvo1983staggered}
Calvo GA (1983{\natexlab{b}}).
\newblock \bibenquote{Staggered prices in a utility-maximizing framework.}
\newblock \emph{Journal of monetary Economics}, \textbf{12}(3), 383--398.

\bibitem[{Carriero \emph{et~al.}(2024)Carriero, Clark, Marcellino, and Mertens}]{carriero2024addressing}
Carriero A, Clark TE, Marcellino M, Mertens E (2024).
\newblock \bibenquote{Addressing COVID-19 outliers in BVARs with stochastic volatility.}
\newblock \emph{Review of Economics and Statistics}, pp. 1--15.

\bibitem[{Cecchetti \emph{et~al.}(2007)Cecchetti, Hooper, Kasman, Schoenholtz, and Watson}]{cecchetti2007understanding}
Cecchetti SG, Hooper P, Kasman BC, Schoenholtz KL, Watson MW (2007).
\newblock \bibenquote{Understanding the evolving inflation process.}
\newblock In \bibenquote{US Monetary Policy Forum,} volume~8, pp. 5--23. University of Chicago Press.

\bibitem[{Chakraborty and Joseph(2017)}]{chakraborty2017machine}
Chakraborty C, Joseph A (2017).
\newblock \bibenquote{Machine learning at central banks.}
\newblock \emph{BoE Working Paper Series No. 674}.

\bibitem[{Chakravarty(2020)}]{chakravarty_2020}
Chakravarty R (2020).
\newblock \bibenquote{The new monetary policy framework: What it means.}
\newblock \emph{Journal of Quantitative Economics}, \textbf{18}, 457--470.

\bibitem[{Challu \emph{et~al.}(2023)Challu, Olivares, Oreshkin, Ramirez, Canseco, and Dubrawski}]{challu2023nhits}
Challu C, Olivares KG, Oreshkin BN, Ramirez FG, Canseco MM, Dubrawski A (2023).
\newblock \bibenquote{N-HITS: Neural hierarchical interpolation for time series forecasting.}
\newblock In \bibenquote{Proceedings of the AAAI Conference on Artificial Intelligence,} volume~37, pp. 6989--6997.

\bibitem[{Chen and Guestrin(2016)}]{chen2016xgboost}
Chen T, Guestrin C (2016).
\newblock \bibenquote{Xgboost: A scalable tree boosting system.}
\newblock In \bibenquote{Proceedings of the 22nd ACM SIGKDD international conference on knowledge discovery and data mining,} pp. 785--794.

\bibitem[{Chinoy \emph{et~al.}(2016)Chinoy, Kumar, and Mishra}]{chinoy2016responsible}
Chinoy SZ, Kumar P, Mishra P (2016).
\newblock \bibenquote{What is responsible for India’s sharp disinflation?}
\newblock In \bibenquote{Disinflation in India,} pp. 425--452. Springer India.

\bibitem[{Cristini and Ferri(2021)}]{cristini2021nonlinear}
Cristini A, Ferri P (2021).
\newblock \bibenquote{Nonlinear models of the Phillips curve.}
\newblock \emph{Journal of Evolutionary Economics}, \textbf{31}(4), 1129--1155.

\bibitem[{Diebold and Mariano(2002)}]{diebold2002comparing}
Diebold FX, Mariano RS (2002).
\newblock \bibenquote{Comparing predictive accuracy.}
\newblock \emph{Journal of Business \& economic statistics}, \textbf{20}(1), 134--144.

\bibitem[{Doser \emph{et~al.}(2023)Doser, Nunes, Rao, and Sheremirov}]{doser2023inflation}
Doser A, Nunes R, Rao N, Sheremirov V (2023).
\newblock \bibenquote{Inflation expectations and nonlinearities in the Phillips curve.}
\newblock \emph{Journal of Applied Econometrics}, \textbf{38}(4), 453--471.

\bibitem[{Doshi-Velez and Kim(2017)}]{doshi2017interpretable}
Doshi-Velez F, Kim B (2017).
\newblock \bibenquote{Towards A Rigorous Science of Interpretable Machine Learning.}
\newblock Retrieved from: \url{https://arxiv.org/abs/1702.08608}.

\bibitem[{Dua(2023)}]{dua_2023}
Dua P (2023).
\newblock \bibenquote{Monetary policy framework in India.}
\newblock In \bibenquote{Macroeconometric Methods: Applications to the Indian Economy,} pp. 39--72. Springer Nature Singapore.

\bibitem[{Dua and Goel(2021)}]{dua2021determinants}
Dua P, Goel D (2021).
\newblock \bibenquote{Determinants of inflation in India.}
\newblock \emph{The Journal of developing areas}, \textbf{55}(2).

\bibitem[{Eichengreen and Gupta(2024)}]{eichengreen2024inflation}
Eichengreen B, Gupta P (2024).
\newblock \bibenquote{Inflation Targeting in India: A Further Assessment.}
\newblock \emph{Margin: The Journal of Applied Economic Research}, \textbf{18}(1-2), 7--42.

\bibitem[{Friedman(2001)}]{friedman2001greedy}
Friedman JH (2001).
\newblock \bibenquote{Greedy function approximation: A gradient boosting machine.}
\newblock \emph{Annals of Statistics}, pp. 1189--1232.

\bibitem[{Friedman(1968)}]{friedman1968role}
Friedman M (1968).
\newblock \bibenquote{The role of monetary policy.}
\newblock \emph{American Economic Review}, \textbf{58}, 1--17.

\bibitem[{Fuhrer(1995)}]{fuhrer1995phillips}
Fuhrer JC (1995).
\newblock \bibenquote{The Phillips curve is alive and well.}
\newblock \emph{New England Economic Review}, pp. 41--57.

\bibitem[{Gali and Gertler(1999)}]{gali1999inflation}
Gali J, Gertler M (1999).
\newblock \bibenquote{Inflation dynamics: A structural econometric analysis.}
\newblock \emph{Journal of Monetary Economics}, \textbf{44}(2), 195--222.

\bibitem[{Ghate and Ahmed(2023)}]{ghate_ahmed_2023}
Ghate C, Ahmed F (2023).
\newblock \bibenquote{On Modernizing Monetary Policy Frameworks in South Asia.}
\newblock In \bibenquote{South Asia's Path to Resilience Growth,} p. 283. International Monetary Fund.

\bibitem[{Giacomini and Rossi(2010)}]{giacomini2010forecast}
Giacomini R, Rossi B (2010).
\newblock \bibenquote{Forecast comparisons in unstable environments.}
\newblock \emph{Journal of Applied Econometrics}, \textbf{25}(4), 595--620.

\bibitem[{Goldstein \emph{et~al.}(2015)Goldstein, Kapelner, Bleich, and Pitkin}]{goldstein2015peeking}
Goldstein A, Kapelner A, Bleich J, Pitkin E (2015).
\newblock \bibenquote{Peeking inside the black box: Visualizing statistical learning with plots of individual conditional expectation.}
\newblock \emph{Journal of Computational and Graphical Statistics}, \textbf{24}(1), 44--65.

\bibitem[{Goulet~Coulombe \emph{et~al.}(2022)Goulet~Coulombe, Leroux, Stevanovic, and Surprenant}]{coulombe2022machine}
Goulet~Coulombe P, Leroux M, Stevanovic D, Surprenant S (2022).
\newblock \bibenquote{How is machine learning useful for macroeconomic forecasting?}
\newblock \emph{Journal of Applied Econometrics}, \textbf{37}(5), 920--964.

\bibitem[{Guidotti \emph{et~al.}(2018)Guidotti, Monreale, Ruggieri, Turini, Giannotti, and Pedreschi}]{guidotti2018survey}
Guidotti R, Monreale A, Ruggieri S, Turini F, Giannotti F, Pedreschi D (2018).
\newblock \bibenquote{A survey of methods for explaining black box models.}
\newblock \emph{ACM computing surveys (CSUR)}, \textbf{51}(5), 1--42.

\bibitem[{Hall \emph{et~al.}(2018)}]{hall2018machine}
Hall AS, \emph{et~al.} (2018).
\newblock \bibenquote{Machine learning approaches to macroeconomic forecasting.}
\newblock \emph{The Federal Reserve Bank of Kansas City Economic Review}, \textbf{103}(63), 2.

\bibitem[{Harvey and Jaeger(1993)}]{harvey1993detrending}
Harvey AC, Jaeger A (1993).
\newblock \bibenquote{Detrending, stylized facts and the business cycle.}
\newblock \emph{Journal of Applied Econometrics}, \textbf{8}(3), 231--247.

\bibitem[{Hazell \emph{et~al.}(2022)Hazell, Herreno, Nakamura, and Steinsson}]{hazell2022slope}
Hazell J, Herreno J, Nakamura E, Steinsson J (2022).
\newblock \bibenquote{The slope of the Phillips Curve: evidence from US states.}
\newblock \emph{The Quarterly Journal of Economics}, \textbf{137}(3), 1299--1344.

\bibitem[{Hochreiter and Schmidhuber(1997)}]{hochreiter1997lstm}
Hochreiter S, Schmidhuber J (1997).
\newblock \bibenquote{Long short-term memory.}
\newblock \emph{Neural Computation}, \textbf{9}(8), 1735--1780.

\bibitem[{Hoerl and Kennard(1970)}]{hoerl1970ridge2}
Hoerl AE, Kennard RW (1970).
\newblock \bibenquote{Ridge regression: Applications to nonorthogonal problems.}
\newblock \emph{Technometrics}, \textbf{12}(1), 69--82.

\bibitem[{Jose \emph{et~al.}(2021)Jose, Shekhar, Kundu, Kishore, and Bhoi}]{jose2021alternative}
Jose J, Shekhar H, Kundu S, Kishore V, Bhoi BB (2021).
\newblock \bibenquote{Alternative Inflation Forecasting Models for India-What Performs Better in Practice?}
\newblock \emph{Reserve Bank of India Occasional Papers}, \textbf{42}(1).

\bibitem[{Joseph(2019)}]{joseph2019parametric}
Joseph A (2019).
\newblock \bibenquote{Parametric inference with universal function approximators.}
\newblock \emph{arXiv preprint arXiv:1903.04209}.

\bibitem[{Kapur(2013)}]{kapur2013phillips}
Kapur M (2013).
\newblock \bibenquote{Revisiting the Phillips curve for India and inflation forecasting.}
\newblock \emph{Journal of Asian Economics}, \textbf{25}, 17--27.

\bibitem[{Kapur and Patra(2000)}]{kapur2000price}
Kapur M, Patra MD (2000).
\newblock \bibenquote{The Price of Low Inflation.}
\newblock \emph{Reserve Bank of India Occasional Papers}, pp. 191--234.

\bibitem[{Keenan(1985)}]{keenan1985tukey}
Keenan DM (1985).
\newblock \bibenquote{A Tukey nonadditivity-type test for time series nonlinearity.}
\newblock \emph{Biometrika}, \textbf{72}(1), 39--44.

\bibitem[{Kishor and Pratap(2023)}]{kishorandpratap2023}
Kishor NK, Pratap B (2023).
\newblock \bibenquote{The Role of Inflation Targeting in Anchoring Long-Run Inflation Expectations: Evidence from India.}
\newblock \emph{Available at SSRN 4633861}.

\bibitem[{Koning \emph{et~al.}(2005)Koning, Franses, Hibon, and Stekler}]{koning2005m3}
Koning AJ, Franses PH, Hibon M, Stekler HO (2005).
\newblock \bibenquote{The M3 competition: Statistical tests of the results.}
\newblock \emph{International journal of forecasting}, \textbf{21}(3), 397--409.

\bibitem[{Kwiatkowski \emph{et~al.}(1992)Kwiatkowski, Phillips, Schmidt, and Shin}]{kwiatkowski1992testing}
Kwiatkowski D, Phillips PC, Schmidt P, Shin Y (1992).
\newblock \bibenquote{Testing the null hypothesis of stationarity against the alternative of a unit root: How sure are we that economic time series have a unit root?}
\newblock \emph{Journal of econometrics}, \textbf{54}(1-3), 159--178.

\bibitem[{Lenza and Primiceri(2020)}]{lenza2020estimate}
Lenza M, Primiceri GE (2020).
\newblock \bibenquote{How to Estimate a VAR after March 2020.}
\newblock \emph{Technical report}, National Bureau of Economic Research.

\bibitem[{Li \emph{et~al.}(2023)Li, Qian, Wang, Long, Chen, and Sun}]{li2023social}
Li Q, Qian T, Wang J, Long R, Chen H, Sun C (2023).
\newblock \bibenquote{Social “win-win” promotion of green housing under the four-subject evolutionary game.}
\newblock \emph{Energy Economics}, \textbf{127}, 107117.

\bibitem[{Lundberg \emph{et~al.}(2019)Lundberg, Erion, and Lee}]{lundberg2019consistent}
Lundberg SM, Erion GG, Lee SI (2019).
\newblock \bibenquote{Consistent Individualized Feature Attribution for Tree Ensembles.}
\newblock Retrieved from: \url{https://arxiv.org/abs/1802.03888}.

\bibitem[{Lundberg and Lee(2017)}]{lundberg2017unified}
Lundberg SM, Lee SI (2017).
\newblock \bibenquote{A Unified Approach to Interpreting Model Predictions.}
\newblock Retrieved from: \url{https://arxiv.org/abs/1705.07874}.

\bibitem[{Malladi(2023)}]{malladi2023benchmark}
Malladi RK (2023).
\newblock \bibenquote{Benchmark Analysis of Machine Learning Methods to Forecast the US Annual Inflation Rate During a High-Decile Inflation Period.}
\newblock \emph{Computational Economics}, pp. 1--41.

\bibitem[{Mazumder(2011)}]{mazumder2011stability}
Mazumder S (2011).
\newblock \bibenquote{The stability of the Phillips curve in India: Does the Lucas critique apply?}
\newblock \emph{Journal of Asian Economics}, pp. 528--539.

\bibitem[{Miller \emph{et~al.}(2017)Miller, Howe, and Sonenberg}]{miller2017explainable}
Miller T, Howe P, Sonenberg L (2017).
\newblock \bibenquote{Explainable AI: Beware of inmates running the asylum or: How I learnt to stop worrying and love the social and behavioural sciences.}
\newblock \emph{arXiv preprint arXiv:1712.00547}.

\bibitem[{Mohanty and John(2015)}]{mohanty2015determinants}
Mohanty D, John J (2015).
\newblock \bibenquote{Determinants of inflation in India.}
\newblock \emph{Journal of Asian Economics}, \textbf{36}, 86--96.

\bibitem[{Molnar \emph{et~al.}(2020)Molnar, Casalicchio, and Bischl}]{molnar2020interpretable}
Molnar C, Casalicchio G, Bischl B (2020).
\newblock \bibenquote{Interpretable machine learning--a brief history, state-of-the-art and challenges.}
\newblock In \bibenquote{Joint European Conference on Machine Learning and Knowledge Discovery in Databases,} pp. 417--431. Springer International Publishing.

\bibitem[{Muduli \emph{et~al.}(2022)Muduli, Nadhanael, and Pattanaik}]{muduli2022assessing}
Muduli S, Nadhanael G, Pattanaik S (2022).
\newblock \bibenquote{Assessing Inflation Expectations Adjusting for Households’ Biases.}
\newblock \emph{RBI Bulletin}, \textbf{76}(12), 97--107.
\newblock Available at SSRN: \url{https://ssrn.com/abstract=4308640}.

\bibitem[{Mullainathan and Spiess(2017)}]{mullainathan2017machine}
Mullainathan S, Spiess J (2017).
\newblock \bibenquote{Machine learning: an applied econometric approach.}
\newblock \emph{Journal of Economic Perspectives}, \textbf{31}(2), 87--106.

\bibitem[{Nakamura(2005)}]{nakamura2005inflation}
Nakamura E (2005).
\newblock \bibenquote{Inflation forecasting using a neural network.}
\newblock \emph{Economics Letters}, \textbf{86}(3), 373--378.

\bibitem[{Ollech and Webel(2023)}]{ollech2023random}
Ollech D, Webel K (2023).
\newblock \bibenquote{A random forest-based approach to combining and ranking seasonality tests.}
\newblock \emph{Journal of Econometric Methods}, \textbf{12}(1), 117--130.

\bibitem[{Oreshkin \emph{et~al.}(2019)Oreshkin, Carpov, Chapados, and Bengio}]{oreshkin2019n}
Oreshkin BN, Carpov D, Chapados N, Bengio Y (2019).
\newblock \bibenquote{N-BEATS: Neural basis expansion analysis for interpretable time series forecasting.}
\newblock \emph{arXiv preprint arXiv:1905.10437}.

\bibitem[{Patra \emph{et~al.}(2021)Patra, Behera, and John}]{patra2021phillips}
Patra MD, Behera H, John J (2021).
\newblock \bibenquote{Is the Phillips Curve in India Dead, Inert and Stirring to Life or Alive and Well?}
\newblock \emph{RBI Bulletin}.

\bibitem[{Patra and Kapur(2012)}]{patra2012monetary}
Patra MD, Kapur M (2012).
\newblock \bibenquote{A monetary policy model for India.}
\newblock \emph{Macroeconomics and Finance in Emerging Market Economies}, \textbf{5}(1), 18--41.

\bibitem[{Patra \emph{et~al.}(2014)Patra, Khundrakpam, and George}]{patra2014post}
Patra MD, Khundrakpam JK, George AT (2014).
\newblock \bibenquote{Post-global crisis inflation dynamics in India: What has changed.}
\newblock In \bibenquote{India Policy Forum,} volume~10, pp. 117--203. National Council of Applied Economic Research.

\bibitem[{Patra and Ray(2010)}]{patra2010inflation}
Patra MD, Ray P (2010).
\newblock \bibenquote{Inflation expectations and monetary policy in India: An empirical exploration.}
\newblock \emph{Technical report}, International Monetary Fund.

\bibitem[{Pattanaik \emph{et~al.}(2020)Pattanaik, Muduli, and Ray}]{pattanaik2020inflation}
Pattanaik S, Muduli S, Ray S (2020).
\newblock \bibenquote{Inflation expectations of households: do they influence wage-price dynamics in India?}
\newblock \emph{Macroeconomics and Finance in Emerging Market Economies}, \textbf{13}(3), 244--263.

\bibitem[{Paul(2009)}]{paul2009phillips}
Paul BP (2009).
\newblock \bibenquote{In search of the Phillips curve for India.}
\newblock \emph{Journal of Asian Economics}, \textbf{20}(4), 479--488.

\bibitem[{Phelps(1967)}]{phelps1967phillips}
Phelps ES (1967).
\newblock \bibenquote{Phillips curves, expectations of inflation and optimal unemployment over time.}
\newblock \emph{Economica}, \textbf{34}(254), 254--281.

\bibitem[{Phillips(1958)}]{phillips1958relation}
Phillips AW (1958).
\newblock \bibenquote{The relation between unemployment and the rate of change of money wage rates in the United Kingdom, 1861-1957.}
\newblock \emph{Economica}, \textbf{25}(100), 283--299.

\bibitem[{Pratap and Sengupta(2019)}]{pratap2019macroeconomic}
Pratap B, Sengupta S (2019).
\newblock \bibenquote{Macroeconomic Forecasting in India: Does Machine Learning Hold the Key to Better Forecasts?}
\newblock \emph{RBI Working Paper Series, No. 04/2019}.

\bibitem[{Roberts(1995)}]{roberts1995new}
Roberts JM (1995).
\newblock \bibenquote{New Keynesian economics and the Phillips curve.}
\newblock \emph{Journal of Money, Credit and Banking}, \textbf{27}(4), 975--984.

\bibitem[{Rossi and Sekhposyan(2016)}]{rossi2016forecast}
Rossi B, Sekhposyan T (2016).
\newblock \bibenquote{Forecast rationality tests in the presence of instabilities, with applications to Federal Reserve and survey forecasts.}
\newblock \emph{Journal of Applied Econometrics}, \textbf{31}(3), 507--532.

\bibitem[{Samuelson and Solow(1960)}]{samuelson1960analytical}
Samuelson PA, Solow RM (1960).
\newblock \bibenquote{Analytical aspects of anti-inflation policy.}
\newblock \emph{The American Economic Review}, \textbf{50}(2), 177--194.

\bibitem[{Sbordone(2002)}]{sbordone2002prices}
Sbordone AM (2002).
\newblock \bibenquote{Prices and unit labor costs: a new test of price stickiness.}
\newblock \emph{Journal of Monetary Economics}, \textbf{49}(2), 265--292.

\bibitem[{Sengupta \emph{et~al.}(2024)Sengupta, Chakraborty, and Singh}]{sengupta2024forecasting}
Sengupta S, Chakraborty T, Singh SK (2024).
\newblock \bibenquote{Forecasting CPI inflation under economic policy and geopolitical uncertainties.}
\newblock \emph{International Journal of Forecasting}.

\bibitem[{Shapley(1953)}]{shapley1953value}
Shapley L (1953).
\newblock \bibenquote{A value for n-person games.}
\newblock \emph{Contributions to the Theory of Games}, \textbf{2}(28), 307--317.

\bibitem[{Singh and Bhoi(2022)}]{singh2022inflation}
Singh N, Bhoi B (2022).
\newblock \bibenquote{Inflation Forecasting in India: Are Machine Learning Techniques Useful?}
\newblock \emph{Reserve Bank of India Occasional Papers}, \textbf{43}(2).

\bibitem[{Srinivasan \emph{et~al.}(2006)Srinivasan, Mahambare, and Ramachandran}]{srinivasan2006modelling}
Srinivasan N, Mahambare V, Ramachandran M (2006).
\newblock \bibenquote{Modelling inflation in India: A critique of the structuralist approach.}
\newblock \emph{Journal of Quantitative Economics}, \textbf{45}.

\bibitem[{Stock and Watson(2007)}]{stock2007why}
Stock JH, Watson MW (2007).
\newblock \bibenquote{Why Has U.S. Inflation Become Harder to Forecast?}
\newblock \emph{Journal of Money, Credit, and Banking}, \textbf{39}(S1), 3--34.

\bibitem[{Stock and Watson(2020)}]{stock2020slack}
Stock JH, Watson MW (2020).
\newblock \bibenquote{Slack and cyclically sensitive inflation.}
\newblock \emph{Journal of Money, Credit and Banking}, \textbf{52}(S2), 393--428.

\bibitem[{Sun \emph{et~al.}(2024)Sun, Xu, and Wang}]{sun2024deep}
Sun C, Xu M, Wang B (2024).
\newblock \bibenquote{Deep learning: spatiotemporal impact of digital economy on energy productivity.}
\newblock \emph{Renewable and Sustainable Energy Reviews}, \textbf{199}, 114501.

\bibitem[{Taylor(1980)}]{taylor1980aggregate}
Taylor JB (1980).
\newblock \bibenquote{Aggregate dynamics and staggered contracts.}
\newblock \emph{Journal of Political Economy}, \textbf{88}(1), 1--23.

\bibitem[{Tibshirani(1996)}]{tibshirani1996lasso}
Tibshirani R (1996).
\newblock \bibenquote{Regression shrinkage and selection via the lasso.}
\newblock \emph{Journal of the Royal Statistical Society Series B: Statistical Methodology}, \textbf{58}(1), 267--288.

\bibitem[{Tsay(1986)}]{tsay1986nonlinearity}
Tsay RS (1986).
\newblock \bibenquote{Nonlinearity tests for time series.}
\newblock \emph{Biometrika}, \textbf{73}(2), 461--466.

\bibitem[{Varian(2014)}]{varian2014big}
Varian HR (2014).
\newblock \bibenquote{Big data: New tricks for econometrics.}
\newblock \emph{Journal of Economic Perspectives}, \textbf{28}(2), 3--28.

\bibitem[{Vilone and Longo(2020)}]{vilone2020explainable}
Vilone G, Longo L (2020).
\newblock \bibenquote{Explainable artificial intelligence: a systematic review.}
\newblock \emph{arXiv preprint arXiv:2006.00093}.

\bibitem[{Vovk \emph{et~al.}(2005)Vovk, Gammerman, and Shafer}]{vovk2005algorithmic}
Vovk V, Gammerman A, Shafer G (2005).
\newblock \emph{Algorithmic learning in a random world}, volume~29.
\newblock Springer.

\bibitem[{Watson(1986)}]{watson1986univariate}
Watson MW (1986).
\newblock \bibenquote{Univariate detrending methods with stochastic trends.}
\newblock \emph{Journal of Monetary Economics}, \textbf{18}(1), 49--75.

\bibitem[{Weisberg and Cook(1982)}]{weisberg1982residuals}
Weisberg S, Cook RD (1982).
\newblock \emph{Residuals and influence in regression}.
\newblock Chapman \& Hall.

\bibitem[{Štrumbelj and Kononenko(2010)}]{strumbelj2010efficient}
Štrumbelj E, Kononenko I (2010).
\newblock \bibenquote{An efficient explanation of individual classifications using game theory.}
\newblock \emph{Journal of Machine Learning Research}, \textbf{11}, 1--18.

\end{thebibliography}

\clearpage
\section*{Ethics Declarations}
The views expressed in the paper are those of authors and do not necessarily reflect the views of the institutions to which they belong.

\subsection*{Funding}
The authors declare that no funds, grants, or other support were received during the preparation of this manuscript.

\subsection*{Competing interests}
The authors have no relevant financial or non-financial interests to disclose. 

\subsection*{Conflict of interest}
The authors have no conflict of interest to declare.






\clearpage

\section*{Appendix A} \label{sec:appendixa}
\setcounter{table}{0}
\renewcommand{\thetable}{A\arabic{table}}
\addcontentsline{toc}{section}{Appendix A}

\subsection*{Model Description}

We employ a comprehensive set of models, encompassing linear, non-linear, and neural network-based approaches, each designed to capture distinct aspects of inflation dynamics.

\textbf{Linear Models:}  
Linear Regression estimates the relationship between a dependent variable \( y \) and predictor variables \( \mathbf{x}_i \) by minimising the residual sum of squares (RSS), defined as:
\begin{equation}
    RSS = \sum_{i=1}^{n} \left( y_i - \beta_0 - \sum_{j=1}^{p} \beta_j x_{ij} \right)^2.
\end{equation}
While effective for linear relationships, this model is sensitive to multicollinearity and outliers.  
Ridge Regression (\cite{hoerl1970ridge2}) addresses these issues by adding an L2 regularisation term, which penalises large coefficients, thereby reducing model variance:
\begin{equation}
    RSS_{\text{ridge}} = \sum_{i=1}^{n} \left( y_i - \beta_0 - \sum_{j=1}^{p} \beta_j x_{ij} \right)^2 + \lambda \sum_{j=1}^{p} \beta_j^2,
\end{equation}
where \(\lambda \geq 0\) is a regularisation parameter controlling the strength of the penalty.  
LASSO Regression (\cite{tibshirani1996lasso}) further enhances model interpretability by performing both regularisation and feature selection. Its L1 penalty encourages sparsity in the coefficients:
\begin{equation}
    RSS_{\text{lasso}} = RSS + \lambda \sum_{j=1}^{p} |\beta_j|.
\end{equation}
Unlike Ridge Regression, LASSO can shrink some coefficients to exactly zero, effectively selecting a subset of predictors.

\textbf{Non-Linear Models:}  
Decision tree-based methods offer flexibility by capturing complex relationships. Random Forest (\cite{breiman2001random}) improves upon individual decision trees by aggregating predictions from \( B \) bootstrapped trees, reducing overfitting:
\begin{equation}
    \hat{f}_{\text{RF}}(\mathbf{x}) = \frac{1}{B} \sum_{b=1}^{B} \hat{f}^{(b)}(\mathbf{x}),
\end{equation}
where each tree \( \hat{f}^{(b)} \) is grown on a random subset of the data and predictors.  
XGBoost (\cite{chen2016xgboost}) extends this concept with gradient boosting, building trees sequentially to correct residual errors:
\begin{equation}
    \hat{y}_i^{(t)} = \hat{y}_i^{(t-1)} + \eta \cdot f_t(\mathbf{x}_i),
\end{equation}
where \( \eta \) is the learning rate. The objective function is augmented with regularisation to prevent overfitting:
\begin{equation}
    \mathcal{L}^{(t)} = \sum_{i=1}^{n} l(y_i, \hat{y}_i^{(t)}) + \Omega(f_t).
\end{equation}

\textbf{Neural Network-Based Models:}  
Neural networks provide powerful frameworks for capturing complex temporal patterns. N-BEATS (\citealp{oreshkin2019n}) is a fully connected architecture that uses interpretable basis functions to model trend and seasonality:
\begin{equation}
    \hat{y}_t = \sum_{i=1}^{K} \mathbf{B}_i \boldsymbol{\theta}_i,
\end{equation}
where \( \mathbf{B}_i \) are basis matrices, and \( \boldsymbol{\theta}_i \) are learned parameters.  
N-HITS (\citealp{challu2023nhits}) enhances N-BEATS by incorporating hierarchical interpolation filters for multi-resolution forecasting:
\begin{equation}
    \hat{y}_t = \sum_{i=1}^{K} \mathbf{H}_i \ast \mathbf{B}_i \boldsymbol{\theta}_i.
\end{equation}
Lastly, BlockRNN with LSTM cells (\citealp{hochreiter1997lstm}) captures long-term dependencies by recursively updating two key components: the hidden state (\( h_t \)), which stores short-term information, and the cell state (\( c_t \)), which retains long-term memory. This update is governed by the function:  

\begin{equation}  
    h_t, c_t = \text{LSTM}(x_t, h_{t-1}, c_{t-1}).  
\end{equation}  

Here, \( x_t \) represents the input at time \( t \), while \( h_{t-1} \) and \( c_{t-1} \) are the hidden and cell states from the previous time step. Additionally, these models can incorporate external variables (\( \mathbf{X}_t \)), enabling them to account for outside factors influencing inflation.

\subsection*{Error metrics employed in the paper}
\begin{itemize}
    \item Median Relative Absolute Error (MdRAE):
MdRAE measures the typical relative error of forecasts compared to a naive forecast. It's robust to outliers, making it useful for datasets with extreme values. MdRAE is scale-independent, allowing for comparisons across different time series.

\begin{equation}
    \text{MdRAE} = \text{median}\left(\frac{|Y_t - \hat{Y}_t|}{|Y_t - Y_{t-1}|}\right)
\end{equation}

    \item Root Mean Squared Error (RMSE):
RMSE quantifies the average magnitude of forecast errors in the same units as the data. It gives higher weight to large errors due to its squared term, making it particularly sensitive to outliers. RMSE is widely used but can be challenging to interpret for non-specialists.

\begin{equation}
    \text{RMSE} = \sqrt{\frac{1}{n}\sum_{t=1}^n (Y_t - \hat{Y}_t)^2}
\end{equation}

    \item Symmetric Mean Absolute Percentage Error (SMAPE):
SMAPE expresses forecast accuracy as a percentage, making it easy to understand across different scales. It treats positive and negative errors symmetrically, addressing some limitations of the traditional MAPE. SMAPE is bounded between 0\% and 200\%, with lower values indicating better accuracy.

\begin{equation}
    \text{SMAPE} = \frac{100\%}{n} \sum_{t=1}^n \frac{|Y_t - \hat{Y}_t|}{(|Y_t| + |\hat{Y}_t|)/2}
\end{equation}

    \item Theil's U:
Theil's U2 statistic, also known as the Theil Inequality Coefficient, is a valuable forecasting accuracy measure that compares the performance of a forecasting model to that of a naive "no change" forecast. The formula for Theil's U2 is given by:

\begin{equation}
    U = \frac{\left[ \frac{1}{n} \sum_{i=1}^{n} \left( Y_i - \hat{Y}_i\right)^2 \right]^{\frac{1}{2}}}{\left[ \frac{1}{n} \sum_{i=1}^{n} Y_i^2 \right]^{\frac{1}{2}} + \left[ \frac{1}{n} \sum_{i=1}^{n} \hat{Y}_i^2 \right]^{\frac{1}{2}}}
\end{equation}

where $Y_i$ represents actual values, $\hat{Y}_i$ represents predicted values, and n is the number of observations. The numerator of this formula is equivalent to the Root Mean Squared Error (RMSE) of the forecast, while the denominator sums the root mean squared values of the actual and predicted series. This statistic is bounded between 0 and 1, with 0 indicating a perfect forecast and 1 signifying the worst possible forecast. Values closer to 0 suggest that the forecasting model outperforms the naive forecast, making Theil's U2 a useful tool for assessing forecast quality and model performance in time series analysis.

\end{itemize}

\begin{table}[!ht]
    \centering
    \caption{1Q-ahead forecasts}
    \scalebox{0.8}{
    \begin{tabular}{l|l|cccc}
    \hline
        \textbf{Specification} & \textbf{Models} & \textbf{MdRAE} & \textbf{RMSE} & \textbf{SMAPE} & \textbf{Theil's U2} \\ \hline
        Backward PC & Linear Regression & 1.54 & 1.77 & 26.54 & 0.17 \\ 
        ~ & Ridge Regression & 1.46 & 1.46 & 22.78 & 0.14 \\ 
        ~ & Lasso Regression & 1.25 & 1.34 & 21.19 & 0.13 \\ 
        ~ & Random Forest & 1.31 & 1.05 & 20.00 & 0.10 \\ 
        ~ & XGBoost & 1.38 & 1.18 & 21.14 & 0.11 \\ 
        ~ & NBEATS & 2.79 & 2.40 & 40.76 & 0.23 \\ 
        ~ & Nhits & 3.68 & 2.38 & 37.86 & 0.20 \\ 
        ~ & BlockRNN & 1.46 & 1.50 & 27.22 & 0.16 \\ 
        \hline
        Forward PC & Linear Regression & 1.64 & 1.89 & 28.69 & 0.18 \\ 
        ~ & Ridge Regression & 1.69 & 1.59 & 26.16 & 0.15 \\ 
        ~ & Lasso Regression & 1.47 & 1.54 & 25.04 & 0.15 \\ 
        ~ & Random Forest & 1.65 & 1.16 & 21.60 & 0.11 \\ 
        ~ & XGBoost & 1.39 & 1.44 & 22.85 & 0.14 \\ 
        ~ & NBEATS & 4.95 & 2.43 & 41.97 & 0.24 \\ 
        ~ & Nhits & 3.14 & 2.37 & 38.97 & 0.20 \\ 
        ~ & BlockRNN & 1.55 & 1.47 & 26.80 & 0.16 \\ 
        \hline
        Hybrid PC & Linear Regression & 1.87 & 1.78 & 27.75 & 0.17 \\ 
        ~ & Ridge Regression & 1.43 & 1.46 & 24.35 & 0.14 \\ 
        ~ & Lasso Regression & 1.45 & 1.34 & 22.44 & 0.13 \\ 
        ~ & Random Forest & 1.32 & 1.06 & 19.84 & 0.10 \\ 
        ~ & XGBoost & 1.04 & 1.29 & 21.12 & 0.12 \\ 
        ~ & NBEATS & 2.57 & 2.98 & 49.87 & 0.27 \\ 
        ~ & Nhits & 2.88 & 2.02 & 33.00 & 0.18 \\ 
        ~ & BlockRNN & 1.78 & 1.72 & 32.85 & 0.19 \\ 
        \hline
        Baseline & Random Walk  & 1.00 & 0.94 & 17.09 & 0.09 \\ \hline
    \end{tabular}}
\end{table}

\begin{table}[!ht]
    \centering
    \caption{2Q-ahead forecasts}
    \scalebox{0.8}{
    \begin{tabular}{l|l|cccc}
    \hline
        \textbf{Specification} & \textbf{Models} & \textbf{MdRAE} & \textbf{RMSE} & \textbf{SMAPE} & \textbf{Theil's U2} \\ \hline
        Backward PC & Linear Regression & 1.20 & 1.87 & 22.99 & 0.17 \\ 
        ~ & Ridge Regression & 1.40 & 1.46 & 22.15 & 0.13 \\ 
        ~ & Lasso Regression & 1.38 & 1.12 & 18.19 & 0.10 \\ 
        ~ & Random Forest & 1.58 & 0.99 & 18.69 & 0.09 \\ 
        ~ & XGBoost & 1.10 & 1.01 & 16.93 & 0.09 \\ 
        ~ & NBEATS & 2.42 & 2.27 & 39.42 & 0.22 \\ 
        ~ & Nhits & 3.37 & 2.07 & 33.14 & 0.17 \\ 
        ~ & BlockRNN & 1.74 & 1.55 & 28.88 & 0.17 \\ 
        \hline
        Forward PC & Linear Regression & 3.57 & 2.45 & 33.99 & 0.22 \\ 
        ~ & Ridge Regression & 2.54 & 1.83 & 30.84 & 0.17 \\ 
        ~ & Lasso Regression & 1.81 & 1.55 & 28.83 & 0.15 \\ 
        ~ & Random Forest & 1.55 & 1.06 & 19.28 & 0.10 \\ 
        ~ & XGBoost & 1.77 & 1.18 & 20.16 & 0.11 \\ 
        ~ & NBEATS & 3.62 & 2.27 & 38.23 & 0.22 \\ 
        ~ & Nhits & 3.53 & 2.19 & 34.45 & 0.18 \\ 
        ~ & BlockRNN & 1.65 & 1.54 & 28.34 & 0.17 \\ 
        \hline
        Hybrid PC & Linear Regression & 2.09 & 2.03 & 28.43 & 0.18 \\ 
        ~ & Ridge Regression & 1.37 & 1.52 & 25.29 & 0.14 \\ 
        ~ & Lasso Regression & 1.12 & 1.24 & 21.02 & 0.12 \\ 
        ~ & Random Forest & 1.43 & 0.93 & 16.72 & 0.09 \\ 
        ~ & XGBoost & 1.93 & 1.05 & 19.35 & 0.10 \\ 
        ~ & NBEATS & 2.99 & 2.41 & 37.81 & 0.22 \\ 
        ~ & Nhits & 2.56 & 1.79 & 26.29 & 0.16 \\ 
        ~ & BlockRNN & 2.37 & 1.80 & 35.92 & 0.20 \\ 
        \hline

        Baseline & Random Walk  & 1.42 & 1.47 & 26.63 & 0.14  \\ \hline
    \end{tabular}}
\end{table}

\begin{table}[!ht]
    \centering
    \caption{3Q-ahead forecasts}
    \scalebox{0.8}{
    \begin{tabular}{l|l|cccc}
    \hline
        \textbf{Specification} & \textbf{Models} & \textbf{MdRAE} & \textbf{RMSE} & \textbf{SMAPE} & \textbf{Theil's U2} \\ \hline
        Backward PC & Linear Regression & 1.48 & 1.59 & 25.44 & 0.14 \\ 
        ~ & Ridge Regression & 1.33 & 1.23 & 19.04 & 0.11 \\ 
        ~ & Lasso Regression & 1.30 & 1.04 & 16.47 & 0.10 \\ 
        ~ & Random Forest & 1.39 & 0.97 & 16.86 & 0.09 \\ 
        ~ & XGBoost & 1.46 & 1.08 & 17.69 & 0.10 \\ 
        ~ & NBEATS & 3.55 & 2.02 & 37.97 & 0.20 \\ 
        ~ & Nhits & 3.75 & 2.25 & 32.05 & 0.19 \\ 
        ~ & BlockRNN & 2.17 & 1.75 & 32.68 & 0.19 \\ 
        \hline
        Forward PC & Linear Regression & 1.43 & 2.14 & 27.76 & 0.19 \\ 
        ~ & Ridge Regression & 1.97 & 1.63 & 26.79 & 0.15 \\ 
        ~ & Lasso Regression & 1.50 & 1.45 & 25.88 & 0.14 \\ 
        ~ & Random Forest & 1.29 & 1.01 & 18.59 & 0.10 \\ 
        ~ & XGBoost & 1.42 & 1.21 & 20.53 & 0.11 \\ 
        ~ & NBEATS & 3.70 & 2.01 & 35.45 & 0.20 \\ 
        ~ & Nhits & 3.65 & 2.01 & 29.88 & 0.17 \\ 
        ~ & BlockRNN & 2.02 & 1.72 & 31.72 & 0.18 \\ 
        \hline
        Hybrid PC & Linear Regression & 1.69 & 1.65 & 26.03 & 0.15 \\ 
        ~ & Ridge Regression & 1.55 & 1.30 & 21.79 & 0.12 \\ 
        ~ & Lasso Regression & 1.36 & 1.12 & 17.92 & 0.10 \\ 
        ~ & Random Forest & 1.27 & 0.94 & 16.84 & 0.09 \\ 
        ~ & XGBoost & 1.55 & 1.04 & 17.77 & 0.10 \\ 
        ~ & NBEATS & 4.92 & 2.65 & 42.10 & 0.23 \\ 
        ~ & Nhits & 2.61 & 1.71 & 27.78 & 0.15 \\ 
        ~ & BlockRNN & 2.82 & 1.97 & 39.19 & 0.22 \\ 
        \hline

        Baseline & Random Walk  & 2.17 & 1.83 & 33.82 & 0.18 \\ \hline
    \end{tabular}}
\end{table}

\begin{table}[!ht]
    \centering
    \caption{4Q-ahead forecasts}
    \scalebox{0.8}{
    \begin{tabular}{l|l|cccc}
    \hline
        \textbf{Specification} & \textbf{Models} & \textbf{MdRAE} & \textbf{RMSE} & \textbf{SMAPE} & \textbf{Theil's U2 } \\ \hline
        Backward PC & Linear Regression & 1.56 & 1.46 & 25.63 & 0.13 \\ 
        ~ & Ridge Regression & 1.43 & 1.16 & 19.38 & 0.10 \\ 
        ~ & Lasso Regression & 1.39 & 1.09 & 17.69 & 0.10 \\ 
        ~ & Random Forest & 1.24 & 0.92 & 16.20 & 0.09 \\ 
        ~ & XGBoost & 1.24 & 1.02 & 16.72 & 0.09 \\ 
        ~ & NBEATS & 2.71 & 2.00 & 38.13 & 0.20 \\ 
        ~ & Nhits & 2.85 & 2.10 & 30.24 & 0.18 \\ 
        ~ & BlockRNN & 1.98 & 1.90 & 34.14 & 0.20 \\ 
        \hline
        Forward PC & Linear Regression & 1.72 & 1.75 & 25.87 & 0.16 \\ 
        ~ & Ridge Regression & 1.47 & 1.45 & 23.18 & 0.13 \\ 
        ~ & Lasso Regression & 1.61 & 1.41 & 22.60 & 0.13 \\ 
        ~ & Random Forest & 1.16 & 0.96 & 16.31 & 0.09 \\ 
        ~ & XGBoost & 1.25 & 1.05 & 16.19 & 0.10 \\ 
        ~ & NBEATS & 2.40 & 1.84 & 33.98 & 0.18 \\ 
        ~ & Nhits & 2.93 & 2.17 & 34.06 & 0.19 \\ 
        ~ & BlockRNN & 1.92 & 1.87 & 33.36 & 0.20 \\ 
        \hline
        Hybrid PC & Linear Regression & 1.27 & 1.43 & 23.75 & 0.13 \\ 
        ~ & Ridge Regression & 1.26 & 1.14 & 17.83 & 0.11 \\ 
        ~ & Lasso Regression & 1.29 & 1.04 & 16.01 & 0.10 \\ 
        ~ & Random Forest & 1.00 & 0.89 & 15.00 & 0.08 \\ 
        ~ & XGBoost & 1.00 & 0.87 & 13.82 & 0.08 \\ 
        ~ & NBEATS & 3.33 & 2.33 & 33.79 & 0.21 \\ 
        ~ & Nhits & 1.82 & 1.63 & 26.59 & 0.14 \\ 
        ~ & BlockRNN & 2.35 & 2.12 & 40.59 & 0.23 \\ 
        \hline

        Baseline & Random Walk  & 1.96 & 2.09 & 39.02 & 0.20 \\ \hline
    \end{tabular}}
\end{table}
\begin{table}[!ht]
    \centering
    \caption{1-quarter ahead forecasts: Using HP Filter for Trend Inflation and Output Gap estimation}
    \scalebox{0.8}{
    \begin{tabular}{l|l|c c c c}
    \hline
        \textbf{Specification} & \textbf{Models} & \textbf{MdRAE} & \textbf{RMSE} & \textbf{SMAPE} & \textbf{Theil's U} \\ \hline
        Backward looking Phillips Curve & Linear Regression & 2.21 & 2.11 & 34.60 & 0.20 \\ 
        ~ & Lasso Regression & 1.69 & 1.65 & 29.12 & 0.16 \\ 
        ~ & Random Forest & \textbf{\textcolor{blue}{1.11}} & \textbf{\textcolor{blue}{1.28}} & 22.94 & \textbf{\textcolor{blue}{0.12}} \\ 
        ~ & Ridge Regression & 2.05 & 1.88 & 31.64 & 0.18 \\ 
        ~ & XGBoost & 1.33 & 1.33 & \textbf{\textcolor{blue}{22.62}} & 0.13 \\ \hline
        Forward looking Phillips Curve & Linear Regression & 1.70 & 2.64 & 51.56 & 0.29 \\ 
        ~ & Lasso Regression & 1.46 & 2.14 & 37.91 & 0.23 \\ 
        ~ & Random Forest & 1.68 & 1.45 & 24.93 & 0.15 \\ 
        ~ & Ridge Regression & 1.45 & 2.36 & 43.58 & 0.25 \\ 
        ~ & XGBoost & 2.00 & 1.51 & 26.00 & 0.15 \\ \hline
        Hybrid Phillips Curve & Linear Regression & 2.48 & 2.12 & 41.57 & 0.21 \\ 
        ~ & Lasso Regression & 1.66 & 1.71 & 31.88 & 0.17 \\ 
        ~ & Random Forest & 1.23 & 1.34 & 23.50 & 0.13 \\ 
        ~ & Ridge Regression & 2.00 & 1.93 & 37.28 & 0.19 \\ 
        ~ & XGBoost & 1.53 & 1.42 & 24.82 & 0.14 \\ \hline
        Baseline & Random Walk  & 1.00 & 0.94 & 17.09 & 0.09 \\ \hline
         \multicolumn{6}{p{\textwidth}}{\small{Note: The table presents 1-quarter ahead forecast error metrics for selected models when trend inflation and output gap is calculated using an Hodrik-Prescott (HP) Filter.}}
    \end{tabular}}
\end{table}

\begin{table}[!ht]
    \centering
    \caption{1-quarter ahead forecasts: Using the first principal component of Supply shocks}
    \scalebox{0.8}{
    \begin{tabular}{l|l|cccc}
\hline
        \textbf{Specification} & \textbf{Models} & \textbf{MdRAE} & \textbf{RMSE} & \textbf{SMAPE} & \textbf{Theil's U} \\ \hline
        Backward looking Phillips Curve & Linear Regression & 2.15 & 1.59 & 29.81 & 0.16 \\ 
        ~ & Lasso Regression & 1.23 & 1.41 & 25.53 & 0.14 \\ 
        ~ & Random Forest & 1.76 & 1.33 & 23.95 & 0.13 \\ 
        ~ & Ridge Regression & 1.76 & 1.46 & 27.34 & 0.14 \\ 
        ~ & XGBoost & 1.70 & 1.64 & 27.64 & 0.16 \\ \hline
        Forward looking Phillips Curve & Linear Regression & 1.95 & 1.76 & 31.90 & 0.18 \\ 
        ~ & Lasso Regression & 1.43 & 1.32 & 24.45 & 0.13 \\ 
        ~ & Random Forest & 1.24 & \textbf{\textcolor{blue}{1.26}} & \textbf{\textcolor{blue}{22.47}} & \textbf{\textcolor{blue}{0.12}} \\ 
        ~ & Ridge Regression & 1.63 & 1.46 & 25.97 & 0.14 \\ 
        ~ & XGBoost & 1.96 & 1.48 & 26.05 & 0.14 \\ \hline
        Hybrid Phillips Curve & Linear Regression & 1.96 & 1.56 & 27.98 & 0.15 \\ 
        ~ & Lasso Regression & 1.43 & 1.35 & 24.82 & 0.13 \\ 
        ~ & Random Forest & \textbf{\textcolor{blue}{1.17}} & 1.29 & 22.94 & 0.13 \\ 
        ~ & Ridge Regression & 1.66 & 1.41 & 26.37 & 0.14 \\ 
        ~ & XGBoost & 1.85 & 1.44 & 26.69 & 0.14 \\ \hline
        Baseline & Random Walk  & 1.00 & 0.94 & 17.09 & 0.09 \\ \hline
        \multicolumn{6}{p{\textwidth}}{\small{Note: The table presents 1-quarter ahead forecast error metrics for selected models when all the supply shocks namely changes in USD/INR exchange rate, crude oil and rainfall deviation, are replaced with the first principal component of the ssupply shocks.}}
    \end{tabular}}
\end{table}

\clearpage
\begin{table}[h]
    \centering
    \caption{1-quarter ahead forecasts: Using the expected inflation measure of \cite{bicchal2019rationality}}
    \scalebox{0.8}{
    \begin{tabular}{l|l|cccc}
    \hline
        Specification & Models & MdRAE & RMSE & SMAPE & Theil's U \\ \hline
        Backward looking Phillips Curve & Linear Regression & 2.35 & 1.96 & 27.38 & 0.16 \\ 
        ~ & Lasso Regression & 1.68 & 1.42 & 18.58 & 0.12 \\
        ~ & Random Forest & 1.03 & 1.14 & 15.27 & 0.09 \\
        ~ & Ridge Regression & 2.66 & 1.61 & 21.56 & 0.13 \\
        ~ & XGBoost & 2.29 & 1.32 & 20.06 & 0.11 \\ \hline
        Forward looking Phillips Curve & Linear Regression & 4.62 & 3.82 & 41.42 & 0.27 \\ 
        ~ & Lasso Regression & 1.05 & \textbf{\textcolor{blue}{1.09}} & \textbf{\textcolor{blue}{13.98}} & \textbf{\textcolor{blue}{0.08}} \\ 
        ~ & Random Forest & 1.48 & 1.34 & 19.08 & 0.11 \\ 
        ~ & Ridge Regression & 3.03 & 1.82 & 26.94 & 0.14 \\ 
        ~ & XGBoost & 1.80 & 1.51 & 18.28 & 0.13 \\ \hline
        Hybrid Phillips Curve & Linear Regression & 3.06 & 2.15 & 30.80 & 0.18 \\ 
        ~ & Lasso Regression & 1.55 & 1.50 & 20.18 & 0.12 \\ 
        ~ & Random Forest & \textbf{\textcolor{blue}{1.02}} & 1.14 & 15.68 & 0.09 \\ 
        ~ & Ridge Regression & 2.13 & 1.63 & 22.68 & 0.13 \\ 
        ~ & XGBoost & 1.06 & 1.15 & 16.12 & 0.10 \\ \hline
        Baseline & Random Walk & 1.00 & 0.76 & 10.03 & 0.06 \\ \hline
        \multicolumn{6}{p{\textwidth}}{\small{Note: The table presents 1-quarter ahead forecast error metrics for selected models. In this case, instead of using the lead of Trend inflation as expected inflation, Google trends based measure of inflation expectations of \cite{bicchal2019rationality} is employed. This measure is available from March quarter of 2003 only therefore, in this case test set is reduced to 12 quarters.}}
    \end{tabular}}
\end{table}
\clearpage

\end{document}